\documentstyle[aps,eqsecnum,epsfig]{revtex}
\def\baselinestretch{1.1}
\begin{document}
\draft
\title{Azimuthal Asymmetries in Heavy Quark Leptoproduction\\
as a Test of pQCD\\}
\author{N.Ya. Ivanov\footnote{e-mail: nikiv@uniphi.yerphi.am}}
\address{Yerevan Physics Institute, Alikhanian Br.2, 375036 Yerevan,
Armenia}
\maketitle

\begin{abstract}
\noindent We analyze the perturbative and parametric stability of the QCD predictions
for the azimuthal asymmetries in heavy quark leptoproduction. At leading order, the
$\cos \varphi $ asymmetry vanishes whereas the  $\cos 2\varphi $ one is of leading twist
and predicted to be about $15\%$ at energies sufficiently above the production threshold.
We calculate the NLO soft-gluon corrections to $\varphi$-dependent leptoproduction
to the next-to-leading logarithmic accuracy.  The soft-gluon approximation provides a good
description of the available exact NLO results at $Q^{2}\lesssim m^{2}$.
Our analysis shows that, contrary to the production cross sections, the $\cos 2\varphi $
asymmetry is practically insensitive to soft radiation for $Q^{2}\lesssim m^{2}$
at energies of the fixed target experiments.
We conclude that the $\cos 2\varphi $ asymmetry is well defined in pQCD: it is stable
both perturbatively and parametrically, and insensitive (in the case of bottom production)
to nonperturbative contributions. Measurements of the azimuthal asymmetries would
provide an excellent test of pQCD applicability to heavy flavor production.
\end{abstract}


\section{Introduction}

In the framework of perturbative QCD, the basic spin-averaged
characteristics of heavy flavor hadro-, photo- and electroproduction are
known exactly up to the next-to-leading order (NLO). During the last ten
years, these NLO results have been widely used for a phenomenological
description of available data (for a review see \cite{1}). At the same time,
the key question remains open: How to test the applicability of QCD at fixed
order to heavy quark production? The problem is twofold. On the one hand,
the NLO corrections are large; they increase the leading order (LO)
predictions for both charm and bottom production cross sections by
approximately a factor of two. For this reason, one could expect that
higher-order corrections, as well as nonperturbative contributions, can be
essential, especially for the $c$-quark case. On the other hand, it is very
difficult to compare pQCD predictions for spin-averaged cross sections with
experimental data directly, without additional assumptions, because of a
high sensitivity of the theoretical calculations to standard uncertainties
in the input QCD parameters. The total uncertainties associated with the
unknown values of the heavy quark mass, $m$, the factorization and
renormalization scales, $\mu _{F}$ and $\mu _{R}$, $\Lambda _{QCD}$ and the
parton distribution functions are so large that one can only estimate the
order of magnitude of the pQCD predictions for total cross sections at fixed
target energies \cite{2,3}.

In recent years, the role of higher-order corrections has been extensively
investigated in the framework of the soft gluon resummation formalism. For a
review see Ref.\cite{4}. Soft gluon (or threshold) resummation is based on
the factorization properties of the cross section near the partonic
threshold and makes it possible to resum to all orders in $\alpha _{s}$ the
leading (Sudakov double) logarithms (LL) and the next-to-leading ones (NLL)
\cite{5,6,7}. Formally resummed cross sections are ill-defined due to the
Landau pole contribution, and a few prescriptions have been proposed to
avoid the renormalon ambiguities \cite{8,9,10}. Unfortunately, numerical
predictions for the heavy quark production cross sections can depend
significantly on the choice of resummation prescription \cite{11}. Another
open question, also closely related to convergence of the perturbative
series, is the role of subleading logarithms which are not, in principle,
under control of the resummation procedure \cite{11,12}.

For this reason, it is of special interest to study those observables that
are well-defined in pQCD. A nontrivial example of such an observable is
proposed in \cite{13,14}, where the single spin asymmetry (SSA) in charm and
bottom production by linearly polarized photons, $\gamma ^{\uparrow
}+N\rightarrow Q+X[\overline{Q}]$, was calculated.\footnote{%
The well-known examples are the shapes of differential cross sections of
heavy flavor production which are sufficiently stable under radiative
corrections.} It was shown that, contrary to the production cross section,
the single spin asymmetry in heavy flavor photoproduction is quantitatively
well defined in pQCD: it is stable, both parametrically and
perturbatively, and insensitive to nonperturbative corrections. Therefore,
measurements of this asymmetry would provide an ideal test of pQCD. As was
shown in Ref. \cite{15}, the SSA in open charm photoproduction can be
measured with an accuracy of about ten percent in the approved E160/E161
experiments at SLAC \cite{16} using the inclusive spectra of secondary
(decay) leptons.

In the present paper we continue the studies of perturbatively stable
observables and calculate the radiative and nonperturbative corrections to
the azimuthal asymmetry (AA) in heavy quark leptoproduction:
\begin{equation}
l(\ell )+N(p)\rightarrow l(\ell -q)+Q(p_{Q})+X[\overline{Q}](p_{X}).
\label{1}
\end{equation}
In the case of unpolarized initial states and neglecting the contribution of $Z-$boson,
the cross section of the reaction (\ref{1}) may be written as
\begin{eqnarray}
\frac{\text{d}\sigma _{lN}}{\text{d}x\text{d}Q^{2}\text{d}\varphi }=\frac{%
\alpha _{em}}{(2\pi )^{2}}\frac{1}{xQ^{2}} &&\Big\{ \left[
1+(1-y)^{2}\right] \sigma _{T}( x,Q^{2}) +2\left(1-y\right) \sigma _{L}(
x,Q^{2})  \nonumber \\
&&+2\left(1-y\right) \sigma _{A}( x,Q^{2}) \cos 2\varphi +(2-y)\sqrt{1-y}%
~\sigma _{I}( x,Q^{2}) \cos \varphi \Big\}.  \label{2}
\end{eqnarray}
The kinematic variables are defined by
\begin{eqnarray}
\bar{S}=\left( \ell +p\right) ^{2},\qquad &Q^{2}=-q^{2},\qquad &x=\frac{Q^{2}}{%
2p\cdot q},  \nonumber \\
y=\frac{p\cdot q}{p\cdot \ell },\qquad &Q^{2}=xy\bar{S},\qquad &\rho =\frac{4m^{2}%
}{\bar{S}}.  \label{3}
\end{eqnarray}
In Eq. (\ref{2}), $\sigma _{T}\,(\sigma _{L})$ is the usual $\gamma ^{*}N$
cross section describing heavy quark production by a transverse
(longitudinal) virtual photon. The third cross section, $\sigma _{A}$, comes
about from interference between transverse states and is responsible for the
SSA which occurs in real photoproduction using linearly polarized photons
\cite{13,14,15}. The fourth cross section, $\sigma _{I}$, originates from
interference between longitudinal and transverse components \cite{16D}.
In the nucleon rest frame, the
azimuth $\varphi $ is the angle between the lepton scattering plane and the
heavy quark production plane, defined by the exchanged photon and the
detected quark $Q$ (see Fig.~\ref{Fg.1}). The covariant definition of
$\varphi $ is
\begin{eqnarray}
\cos \varphi &=&\frac{r\cdot n}{\sqrt{-r^{2}}\sqrt{-n^{2}}},\qquad \qquad
\sin \varphi =\frac{Q^{2}\sqrt{1/x^{2}+4m_{N}^{2}/Q^{2}}}{2\sqrt{-r^{2}}%
\sqrt{-n^{2}}}~n\cdot \ell ,  \label{4} \\
r^{\mu } &=&\varepsilon ^{\mu \nu \alpha \beta }p_{\nu }q_{\alpha }\ell
_{\beta },\qquad \qquad \quad n^{\mu }=\varepsilon ^{\mu \nu \alpha \beta
}p_{\nu }q_{\alpha }p_{Q\beta }.  \label{5}
\end{eqnarray}
In Eqs. (\ref{3}) and (\ref{4}), $m$ and $m_{N}$ are the masses of the heavy
quark and the target, respectively.

In leading order pQCD, the $\cos \varphi $ dependence vanishes,
 $\sigma _{I}^{{\rm Born}}(x,Q^{2})=0$. For this reason, in this paper we
restrict ourselves to the azimuthal asymmetry, $A(\rho ,x,Q^{2})$,
associated with the $\cos 2\varphi $ distribution:
\begin{equation}
A(\rho ,x,Q^{2})=\frac{\text{d}^{3}\sigma _{lN}(\varphi =0)+\text{d}%
^{3}\sigma _{lN}(\varphi =\pi )-2\text{d}^{3}\sigma _{lN}(\varphi =\pi /2)}{%
\text{d}^{3}\sigma _{lN}(\varphi =0)+\text{d}^{3}\sigma _{lN}(\varphi =\pi
)+2\text{d}^{3}\sigma _{lN}(\varphi =\pi /2)}=\frac{\varepsilon \,\sigma
_{A}( x,Q^{2}) }{\sigma _{T}( x,Q^{2}) +\varepsilon
\,\sigma _{L}( x,Q^{2}) },  \label{6}
\end{equation}
where $\varepsilon ={{{{%
{\displaystyle {2(1-y) \over 1+(1-y)^{2}}}%
}}}}$ and $\text{d}^{3}\sigma _{lN}(\varphi )\equiv {%
{\displaystyle {\text{d}^{3}\sigma _{lN} \over \text{d}x\text{d}Q^{2}\text{d}\varphi }}%
}( \rho ,x,Q^{2},\varphi) $. Note that the asymmetry defined by Eq. (\ref{6})
is simply related to the mean value of $\cos 2\varphi $:
\begin{figure}
\begin{center}
\mbox{\epsfig{file=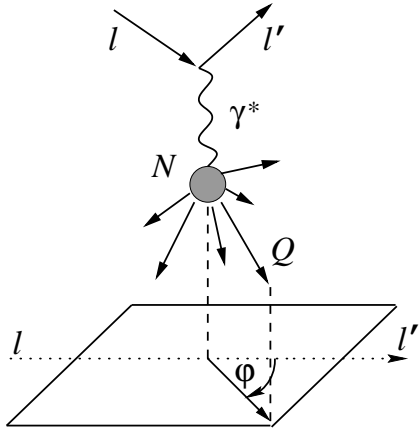,width=200pt}}
\caption{\label{Fg.1}\small Definition of the azimuthal angle $\varphi$ in the nucleon rest frame.}
\end{center}
\end{figure}
\begin{equation}
A(\rho ,x,Q^{2})=2\langle \cos 2\varphi \rangle (\rho ,x,Q^{2}),\qquad
\qquad \langle \cos 2\varphi \rangle (\rho ,x,Q^{2})=
\frac{\int\limits_{0}^{2\pi }\text{d}\varphi \cos 2\varphi%
{\displaystyle {\text{d}^{3}\sigma _{lN} \over \text{d}x\text{d}Q^{2}\text{d}\varphi }}%
( \rho ,x,Q^{2},\varphi ) }{\int\limits_{0}^{2\pi }\text{d}\varphi
{\displaystyle {\text{d}^{3}\sigma _{lN} \over \text{d}x\text{d}Q^{2}\text{d}\varphi }}%
( \rho ,x,Q^{2},\varphi ) }.  \label{8}
\end{equation}

In this paper, we calculate the NLO corrections to the $\cos 2\varphi$ asymmetry
to the next-to-leading logarithmic accuracy (so-called soft-gluon
approximation). Also we analyze the nonperturbative contributions to the AA
due to the gluon transverse motion in the target. Our main results can be
formulated as follows:
\begin{itemize}
\item  The azimuthal asymmetry defined by Eq. (\ref{6}) is of leading twist; at
energies sufficiently above the production threshold, it is predicted to be
about $15\%$ for both charm and bottom quark production.
\item  The soft-gluon approximation provides a good description of the available
exact NLO results on leptoproduction in the region of relatively low virtualities,
$Q^{2}\lesssim m^{2}$;  when $Q^{2}\gg m^{2}$, the quality of the NLL
approximation becomes worse.
\item  Contrary to the production cross sections, the $\cos 2\varphi$ asymmetry
in azimuthal distributions of both charm and bottom quark is practically
insensitive to radiative corrections at $Q^{2}\lesssim m^{2}$. This implies that
large soft-gluon contributions to the $\varphi$-dependent and $\varphi$-integrated
cross sections cancel each other in Eq. (\ref{6}) with a good accuracy.
\item  pQCD predictions for the $\cos 2\varphi$ asymmetry are parametrically
stable; to within few percent, they are insensitive to standard uncertainties in
the QCD input parameters: $\mu _{R}$, $\mu _{F}$, $\Lambda _{QCD}$ and
in the gluon distribution function.
\item  Nonperturbative corrections to the $b$-quark azimuthal asymmetry due
to the gluon transverse motion in the target are negligible. Because of the
smallness of the $c$-quark mass, the analogous corrections to $A(\rho
,x,Q^{2})$ in the charm case are larger; they are of the order of 20\% at $%
Q^{2}\lesssim m^{2}$.
\end{itemize}

We conclude that, in contrast with the production cross sections, the
$\cos 2\varphi$ asymmetry in heavy quark leptoproduction, $A(\rho ,x,Q^{2})$,
is an observable quantitatively well defined in pQCD: it is stable, both
parametrically and perturbatively, and insensitive (in the case of bottom
production) to nonperturbative corrections. Measurements of the AA in
bottom production would provide an ideal test of the conventional parton
model based on pQCD.

Concerning the experimental aspects, azimuthal asymmetries in bottom
production can, in principle, be measured at HERA using the angular distributions
of secondary (decay) leptons \cite{15}. AAs in charm leptoproduction can also be
measured in the COMPASS and HERMES experiments. Due to the relatively
low $c$-quark mass, data on the $D$-meson azimuthal distributions would make
it possible to clarify the role of subleading twist contributions.

The paper is organized as follows. In Section II we analyze the LO and NLO
parton level predictions for $\varphi $-dependent leptoproduction of heavy
flavor in the single-particle inclusive kinematics. We check the quality of
the soft-gluon approach against available exact results and discuss the
region of applicability of the NLL approximation. Hadron level predictions
for $A(\rho ,x,Q^{2})$ are given  in Section III. We consider in detail the
pQCD contributions and nonperturbative corrections to the $\cos 2\varphi$
asymmetry at the HERMES, SLAC, COMPASS and HERA energies.

\section{ Partonic Cross Sections}

\subsection{Born level predictions}

At leading order, ${\cal O}(\alpha _{em}\alpha _{s})$, the only partonic
subprocess which is responsible for heavy quark leptoproduction is the
two-body photon-gluon fusion:
\begin{equation}
\gamma ^{*}(q)+g(k_{g})\rightarrow Q(p_{Q})+\overline{Q}(p_{\stackrel{\_}{Q}%
}).  \label{10}
\end{equation}
The $\gamma ^{*}g$ cross sections, $\hat{\sigma}_{k}^{{\rm Born}}$ ($%
k=T,L,A,I$), corresponding to the Born diagrams are \cite{17a,17}:
\begin{eqnarray}
\hat{\sigma}_{T}^{{\rm Born}}( \hat{\rho},\hat{x}) &=&\frac{\pi
e_{Q}^{2}\alpha _{em}\alpha _{s}}{2m^{2}}\hat{\rho}\Big\{\left[ (1-\hat{x}%
)^{2}+\hat{x}^{2}+\hat{\rho}(1-\hat{x}-\hat{\rho}/2)\right] \ln \frac{%
1+\beta }{1-\beta }  \nonumber \\
&&\qquad \qquad \quad {}-\left[ (1-\hat{x})^{2}-\hat{x}(2-3\hat{x})+(1-\hat{x%
})\hat{\rho}\right] \beta \Big\},  \nonumber \\
\hat{\sigma}_{L}^{{\rm Born}}( \hat{\rho},\hat{x}) &=&\frac{\pi
e_{Q}^{2}\alpha _{em}\alpha _{s}}{m^{2}}\hat{\rho}\hat{x}\Big\{ -\hat{\rho}%
\ln \frac{1+\beta }{1-\beta }+2(1-\hat{x})\beta \Big\},  \label{11} \\
\hat{\sigma}_{A}^{{\rm Born}}( \hat{\rho},\hat{x}) &=&\frac{\pi
e_{Q}^{2}\alpha _{em}\alpha _{s}}{2m^{2}}\hat{\rho}^{2}\Big\{ (1-2\hat{x}-%
\hat{\rho}/2)\ln \frac{1+\beta }{1-\beta }-(1-\hat{x})(1-2\hat{x}/\hat{\rho}%
)\beta \Big\},  \nonumber \\
\hat{\sigma}_{I}^{{\rm Born}}( \hat{\rho},\hat{x}) &=&0.
\nonumber
\end{eqnarray}
In Eqs. (\ref{11}), $e_{Q}$ is the quark charge in units of electromagnetic
coupling constant and we use the following definition of partonic kinematic
variables:
\begin{eqnarray}
s=\left( q+k_{g}\right) ^{2},\qquad \qquad &&\hat{x}=\frac{Q^{2}}{2q\cdot
k_{g}},  \nonumber \\
\beta =\sqrt{1-\frac{4m^{2}}{s}},\qquad \qquad &&\hat{\rho}=\frac{4m^{2}}{%
s+Q^{2}}.  \label{12}
\end{eqnarray}

Note that the $\cos \varphi $ dependence vanishes due to the $%
Q\leftrightarrow \overline{Q}$ symmetry which, at leading order, requires
invariance under $\varphi \rightarrow \varphi +\pi $ \cite{18}.

The hadron level cross sections, $\sigma _{k}( x,Q^{2}) $ ($%
k=T,L,A,I$), have the form
\begin{equation}
\sigma _{k}( x,\xi ) =\int\limits_{x+4x/\xi }^{1}\text{d}%
z\,g(z,\mu _{F})\,\hat{\sigma}_{k}\left( \frac{4x}{z\xi },\frac{x}{z}\right)
,\qquad \qquad k_{g}=zp,\qquad \qquad \xi =\frac{Q^{2}}{m^{2}},  \label{13}
\end{equation}
where $g(z,\mu _{F})$ describes gluon density in a nucleon $N$ evaluated at
a factorization scale $\mu _{F}$. The partonic cross sections, $\hat{\sigma}_{k}$,
are functions of $\hat{\rho}$ and $\hat{x}$ defined by Eq. (\ref{12}).

\subsection{Soft-gluon corrections at NLO}

To take into account the NLO contributions, one needs to calculate the
virtual ${\cal O}(\alpha _{em}\alpha _{s}^{2})$ corrections to the Born
process (\ref{10}) and the real gluon emission:
\begin{equation}
\gamma ^{*}(q)+g(k_{g})\rightarrow Q(p_{Q})+\overline{Q}(p_{\stackrel{\_}{Q}%
})+g(p_{g}).  \label{15}
\end{equation}
The partonic invariants describing the single-particle inclusive (1PI) kinematics
are
\begin{eqnarray}
s^{\prime }=2q\cdot k_{g}=s+Q^{2}=zS^{\prime },\qquad \qquad &&t_{1}=\left(
k_{g}-p_{Q}\right) ^{2}-m^{2}=zT_{1},  \nonumber \\
s_{4}=s^{\prime }+t_{1}+u_{1},\qquad \qquad &&u_{1}=\left( q-p_{Q}\right)
^{2}-m^{2}=U_{1},  \label{16}
\end{eqnarray}
where $s_{4}$ measures the inelasticity of the reaction (\ref{15}). The
corresponding 1PI hadron level variables describing the reaction (\ref{1}) are
\begin{eqnarray}
S^{\prime }=2q\cdot p=S+Q^{2},\qquad \qquad &&T_{1}=\left( p-p_{Q}\right)
^{2}-m^{2},  \nonumber \\
S_{4}=S^{\prime }+T_{1}+U_{1},\qquad \qquad &&U_{1}=\left( q-p_{Q}\right)
^{2}-m^{2}.  \label{17}
\end{eqnarray}
We neglect the photon-(anti)quark fusion subprocesses. This is justified as
their contributions vanish at LO and are small at NLO \cite{19}.

The exact NLO calculations of the unpolarized heavy quark production in $%
\gamma g$ \cite{20,21}, $\gamma ^{*}g$ \cite{19}, and $gg$ \cite{22,23}
collisions show that, near the partonic threshold, a strong logarithmic
enhancement of the cross sections takes place in the collinear, $\vec{p}%
_{g,T} $ $\rightarrow 0$, and soft, $\vec{p}_{g}\rightarrow 0$, limits. This
threshold (or soft-gluon) enhancement has universal nature in the
perturbation theory and originates from incomplete cancellation of the soft
and collinear singularities between the loop and the bremsstrahlung
contributions. Large leading and next-to-leading threshold logarithms can be
resummed to all orders of perturbative expansion using the appropriate
evolution equations \cite{5,6,7}. The analytic results for the resummed
cross sections are ill-defined due to the Landau pole in the coupling
strength $\alpha _{s}$. However, if one considers the obtained expressions
as generating functionals of the perturbative theory and re-expands them at
fixed order in $\alpha _{s}$, no divergences associated with the Landau pole
are encountered.

Soft-gluon resummation for the photon-gluon fusion has been performed in Ref.%
\cite{24} and checked in Refs.\cite{25,14}. To NLL accuracy, the
perturbative expansion for the partonic cross sections, d$^{2}\hat{\sigma}%
_{k}/$d$t_{1}$d$u_{1}$ ($k=T,L,A,I$), can be written in a factorized form as
\begin{equation}
s^{\prime 2}\frac{\text{d}^{2}\hat{\sigma}_{k}}{\text{d}t_{1}\text{d}u_{1}}%
( s^{\prime },t_{1},u_{1}) =B_{k}^{\text{{\rm Born}}}(
s^{\prime },t_{1},u_{1}) \left\{ \delta ( s^{\prime
}+t_{1}+u_{1}) +\sum_{n=1}^{\infty }\left( \frac{\alpha _{s}C_{A}}{\pi
}\right) ^{n}K^{(n)}( s^{\prime },t_{1},u_{1}) \right\} ,
\label{18}
\end{equation}
with the Born level distributions $B_{k}^{\text{{\rm Born}}}$ given by
\begin{eqnarray}
B_{T}^{\text{{\rm Born}}}( s^{\prime },t_{1},u_{1}) &=&\pi
e_{Q}^{2}\alpha _{em}\alpha _{s}\left[ \frac{t_{1}}{u_{1}}+\frac{u_{1}}{t_{1}%
}+4\left( \frac{s}{s^{\prime }}-\frac{m^{2}s^{\prime }}{t_{1}u_{1}}\right)
\left( \frac{s^{\prime }(m^{2}-Q^{2}/2)}{t_{1}u_{1}}+\frac{Q^{2}}{s^{\prime }%
}\right) \right] ,  \nonumber \\
B_{L}^{\text{{\rm Born}}}( s^{\prime },t_{1},u_{1}) &=&\pi
e_{Q}^{2}\alpha _{em}\alpha _{s}\left[ \frac{8Q^{2}}{s^{\prime }}\left(
\frac{s}{s^{\prime }}-\frac{m^{2}s^{\prime }}{t_{1}u_{1}}\right) \right] ,
\nonumber \\
B_{A}^{\text{{\rm Born}}}( s^{\prime },t_{1},u_{1}) &=&\pi
e_{Q}^{2}\alpha _{em}\alpha _{s}\left[ 4\left( \frac{s}{s^{\prime }}-\frac{%
m^{2}s^{\prime }}{t_{1}u_{1}}\right) \left( \frac{m^{2}s^{\prime }}{%
t_{1}u_{1}}+\frac{Q^{2}}{s^{\prime }}\right) \right] ,  \label{19} \\
B_{I}^{\text{{\rm Born}}}( s^{\prime },t_{1},u_{1}) &=&\pi
e_{Q}^{2}\alpha _{em}\alpha _{s}\left[ 4\sqrt{Q^{2}}\left( \frac{t_{1}u_{1}s%
}{s^{\prime 2}}-m^{2}\right) ^{1/2}\frac{t_{1}-u_{1}}{t_{1}u_{1}}\left( 1-%
\frac{2Q^{2}}{s^{\prime }}-\frac{2m^{2}s^{\prime }}{t_{1}u_{1}}\right)
\right] .  \nonumber
\end{eqnarray}

Note that the functions $K^{(n)}( s^{\prime },t_{1},u_{1}) $ in
Eq. (\ref{18}) originate from the collinear and soft limits. Since the
azimuthal angle $\varphi $ is the same for both $\gamma ^{*}g$ and $Q%
\overline{Q}$ center-of-mass systems in these limits, the functions $%
K^{(n)}( s^{\prime },t_{1},u_{1}) $ are also the same for all $%
k=T,L,A,I$. At NLO, the soft-gluon corrections to NLL accuracy in the $%
\overline{\text{MS}}$ scheme are
\begin{eqnarray}
K^{(1)}( s^{\prime },t_{1},u_{1}) &=&2\left[ \frac{\ln \left(
s_{4}/m^{2}\right) }{s_{4}}\right] _{+}-\left[ \frac{1}{s_{4}}\right]
_{+}\left\{ 1+\ln \left( \frac{u_{1}}{t_{1}}\right) -\left( 1-\frac{2C_{F}}{%
C_{A}}\right) \left( 1+\text{Re}L_{\beta }\right) +\ln \left( \frac{\mu ^{2}%
}{m^{2}}\right) \right\}  \nonumber \\
&&+\delta ( s_{4}) \ln \left( \frac{-u_{1}}{m^{2}}\right) \ln
\left( \frac{\mu ^{2}}{m^{2}}\right) ,  \label{20}
\end{eqnarray}
where we use $\mu =\mu _{F}=\mu _{R}$. In Eq. (\ref{20}), $C_{A}=N_{c}$ and $%
C_{F}=(N_{c}^{2}-1)/(2N_{c})$, where $N_{c}$ is number of colors, while $%
L_{\beta }=(1-2m^{2}/s)\{\ln [(1-\beta )/(1+\beta )]+$i$\pi \}$. The
single-particle inclusive ''plus`` distributions are defined by
\begin{equation}
\left[ \frac{\ln ^{l}\left( s_{4}/m^{2}\right) }{s_{4}}\right]
_{+}=\lim_{\epsilon \rightarrow 0}\left\{ \frac{\ln ^{l}\left(
s_{4}/m^{2}\right) }{s_{4}}\theta ( s_{4}-\epsilon ) +\frac{1}{l+1%
}\ln ^{l+1}\left( \frac{\epsilon }{m^{2}}\right) \delta ( s_{4})
\right\} .  \label{21}
\end{equation}
For any sufficiently regular test function $h(s_{4})$, Eq. (\ref{21}) gives
\begin{equation}
\int\limits_{0}^{s_{4}^{\max }}\text{d}s_{4}\,h(s_{4})\left[ \frac{\ln
^{l}\left( s_{4}/m^{2}\right) }{s_{4}}\right]
_{+}=\int\limits_{0}^{s_{4}^{\max }}\text{d}s_{4}\left[ h(s_{4})-h(0)\right]
\frac{\ln ^{l}\left( s_{4}/m^{2}\right) }{s_{4}}+\frac{1}{l+1}h(0)\ln
^{l+1}\left( s_{4}^{\max }/m^{2}\right) .  \label{22}
\end{equation}

In Eq. (\ref{20}) , we have preserved the NLL terms for the scale-dependent
logarithms too. We have checked that the results (\ref{19}) and (\ref{20})
agree to NLL accuracy with the exact ${\cal O}(\alpha _{em}\alpha _{s}^{2})$
calculations of the photon-gluon cross sections $\hat{\sigma}_{T}$ and $\hat{%
\sigma}_{L}$ given in Ref.\cite{19}.

To perform a numerical investigation of the results (\ref{19}) and (\ref{20}%
), it is convenient to introduce for the fully inclusive (integrated over $%
t_{1}$ and $u_{1}$) cross sections, $\hat{\sigma}_{k}$ ($k=T,L,A,I$),{\large %
\ }the dimensionless coefficient functions $c_{k}^{(n,l)}$,
\begin{equation}
\hat{\sigma}_{k}(\eta ,\xi ,\mu ^{2})=\frac{e_{Q}^{2}\alpha _{em}\alpha
_{s}(\mu ^{2})}{m^{2}}\sum_{n=0}^{\infty }\left( 4\pi \alpha _{s}(\mu
^{2})\right) ^{n}\sum_{l=0}^{n}c_{k}^{(n,l)}(\eta ,\xi )\ln ^{l}\left( \frac{%
\mu ^{2}}{m^{2}}\right) ,  \label{23}
\end{equation}
where the variable $\eta $ measures the distance to the partonic threshold:
\begin{equation}
\eta =\frac{s}{4m^{2}}-1,\qquad \qquad \xi =\frac{Q^{2}}{m^{2}}.  \label{23a}
\end{equation}

Concerning the NLO scale-independent coefficient functions, only $%
c_{T}^{(1,0)}$ and $c_{L}^{(1,0)}$ are known exactly \cite{19,26}. As to the $%
\mu $-dependent coefficients, they can by calculated explicitly using the
renormalization group equation:
\begin{equation}
\frac{\text{d}\hat{\sigma}_{k}(s^{\prime },Q^{2},\mu ^{2})}{\text{d}\ln \mu
^{2}}=-\int\limits_{z_{\min }}^{1}\text{d}z\,\hat{\sigma}_{k}(zs^{\prime
},Q^{2},\mu ^{2})P_{gg}(z),  \label{24}
\end{equation}
where $z_{\min }=(4m^{2}+Q^{2})/s^{\prime }$, $\hat{\sigma}_{k}(s^{\prime
},Q^{2},\mu )$ are the cross sections resummed to all orders in $\alpha _{s}$
and $P_{gg}(z)$ is the corresponding (resummed) Altarelli-Parisi gluon-gluon
splitting function. Expanding Eq. (\ref{24}) in $\alpha _{s}$, one can find
\cite{24,14}
\begin{equation}
c_{k}^{(1,1)}(s^{\prime },\xi )=\frac{1}{4\pi ^{2}}\int\limits_{z_{\min
}}^{1}\text{d}z\left[ b_{2}\delta (1-z)-\,P_{gg}^{(0)}(z)\right]
c_{k}^{(0,0)}(zs^{\prime },\xi ),  \label{25}
\end{equation}
where $b_{2}=(11C_{A}-2n_{f})/12$ is the first coefficient of the
$\beta( \alpha _{s}) $-function expansion and $n_{f}$ is the number
of active quark flavors. The one-loop gluon splitting function is \cite{27}:
\begin{equation}
P_{gg}^{(0)}(z)=\lim_{\epsilon \rightarrow 0}\left\{ \left( \frac{z}{1-z}+%
\frac{1-z}{z}+z(1-z)\right) \theta ( 1-z-\epsilon ) +\delta
(1-z)\ln \epsilon \right\} C_{A}+b_{2}\delta (1-z).  \label{26}
\end{equation}

\begin{figure}
\begin{center}
\begin{tabular}{cc}
\mbox{\epsfig{file=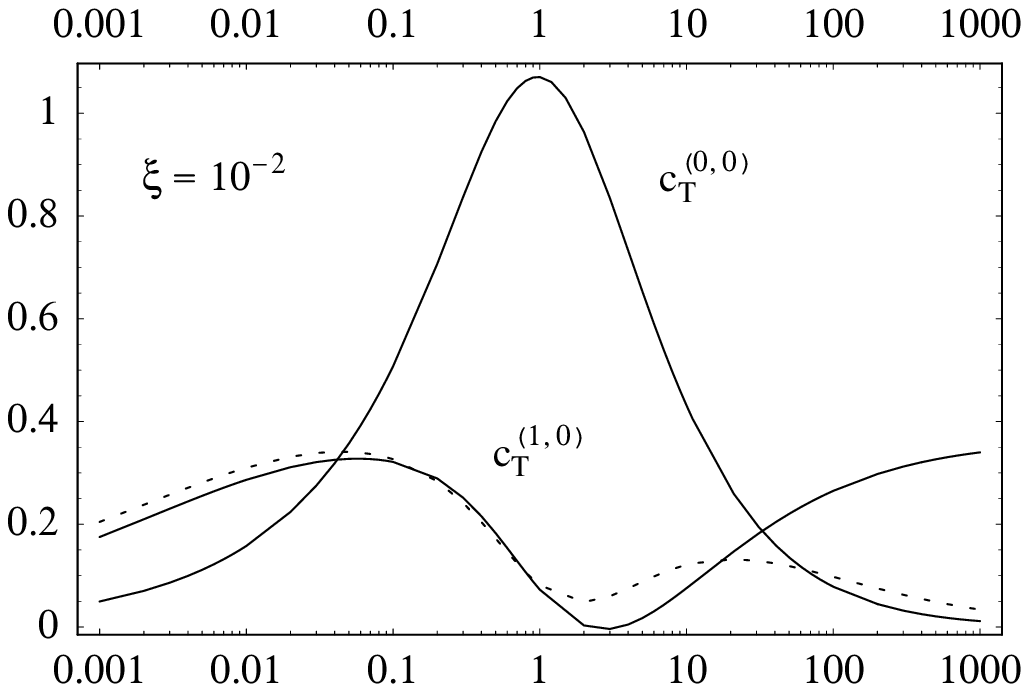,width=213pt}}&
\mbox{\epsfig{file=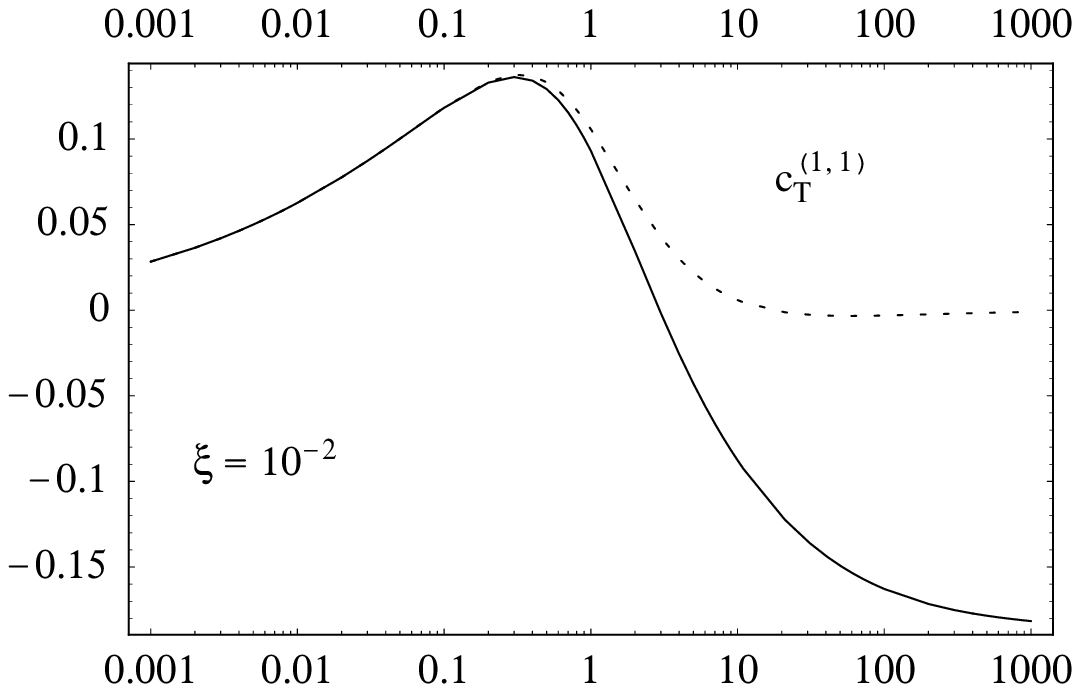,width=213pt}}\\
\mbox{\epsfig{file=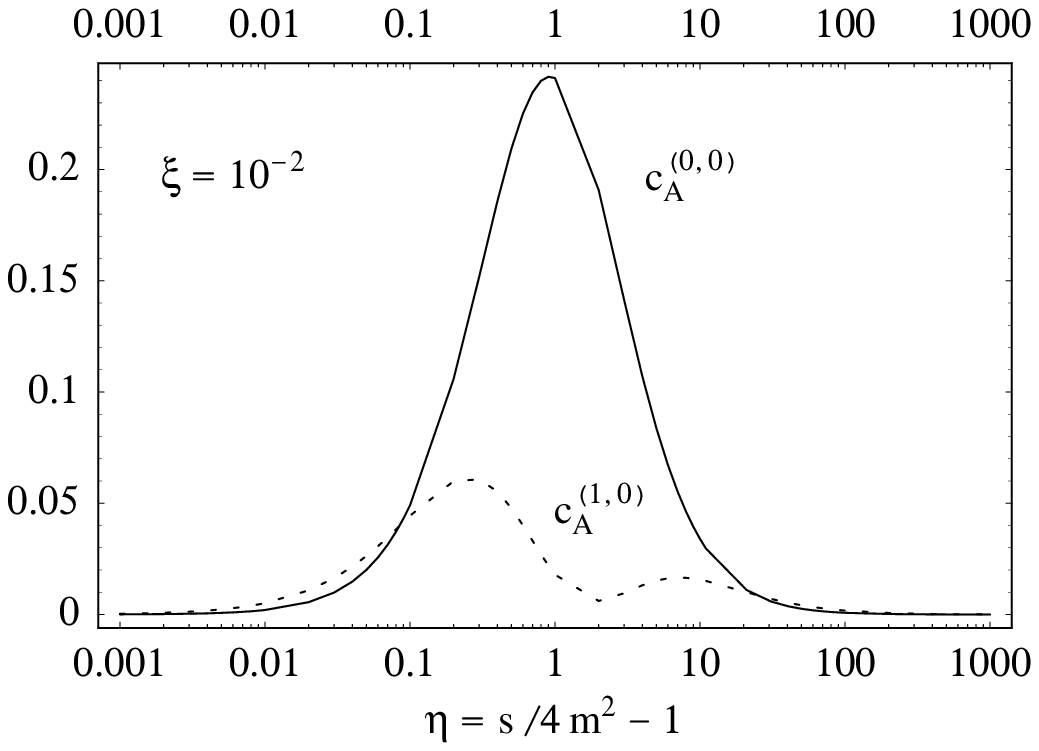,width=213pt}}&
\mbox{\epsfig{file=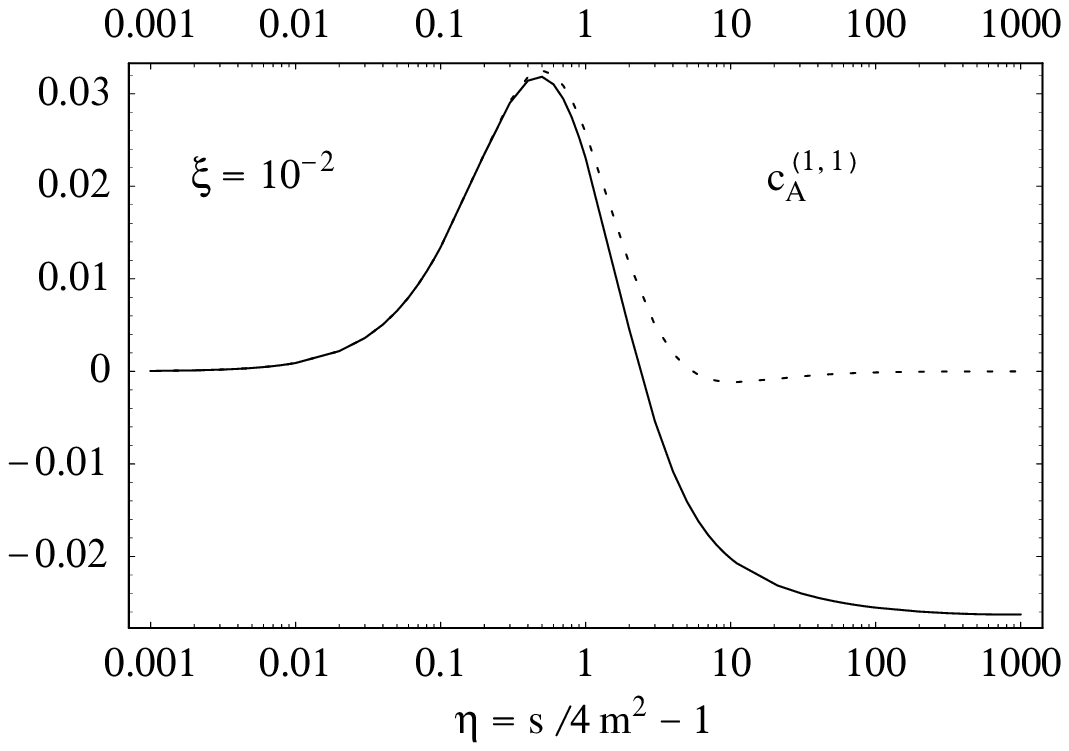,width=213pt}}\\
\end{tabular}
\caption{\label{Fg.3}\small $c_{T}^{(k,l)}(\eta,\xi )$ and
$c_{A}^{(k,l)}(\eta,\xi )$ coefficient functions at $\xi =10^{-2}$. Plotted
are the available exact results (solid lines) and the NLL approximation
(dotted lines).} 
\end{center}
\end{figure}

\begin{figure}
\begin{center}
\begin{tabular}{cc}
\mbox{\epsfig{file=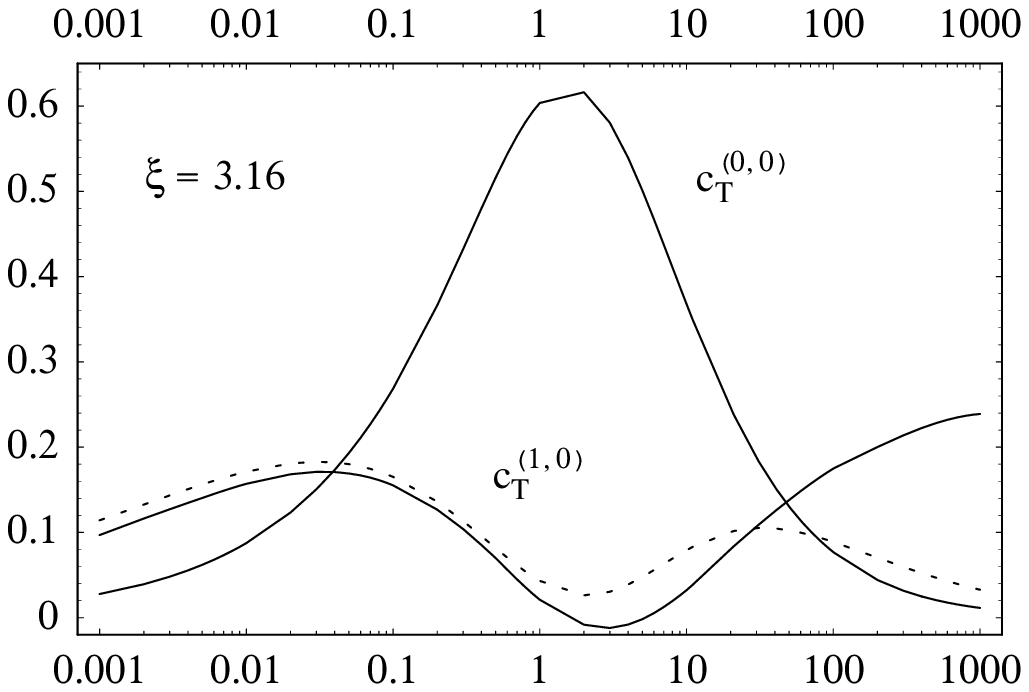,width=213pt}}&
\mbox{\epsfig{file=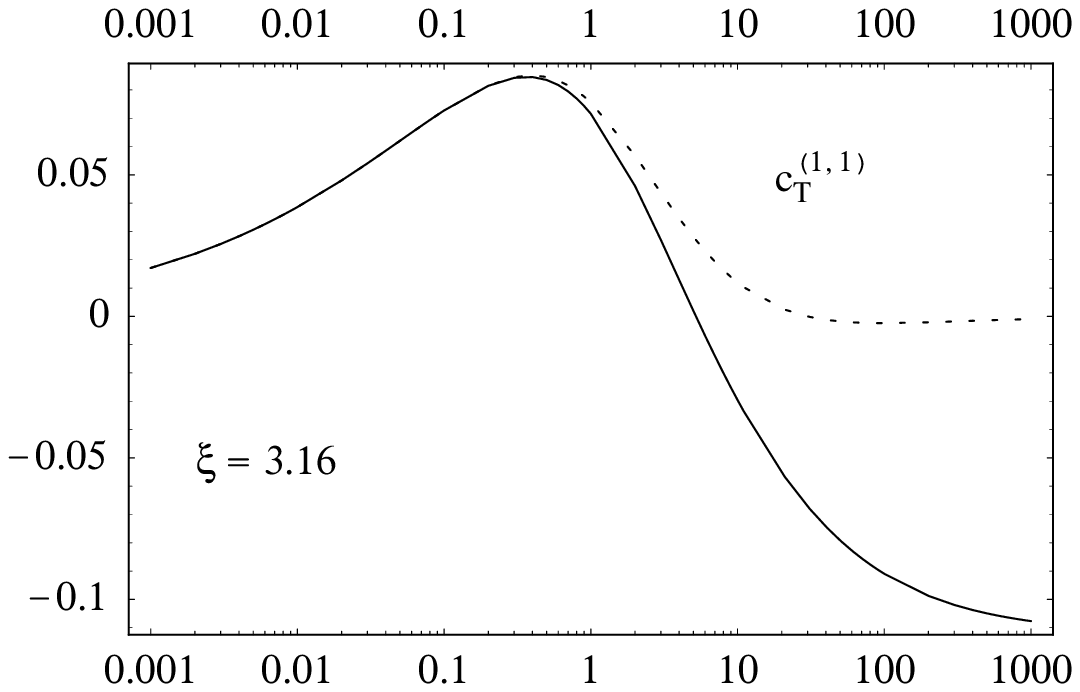,width=213pt}}\\
\mbox{\epsfig{file=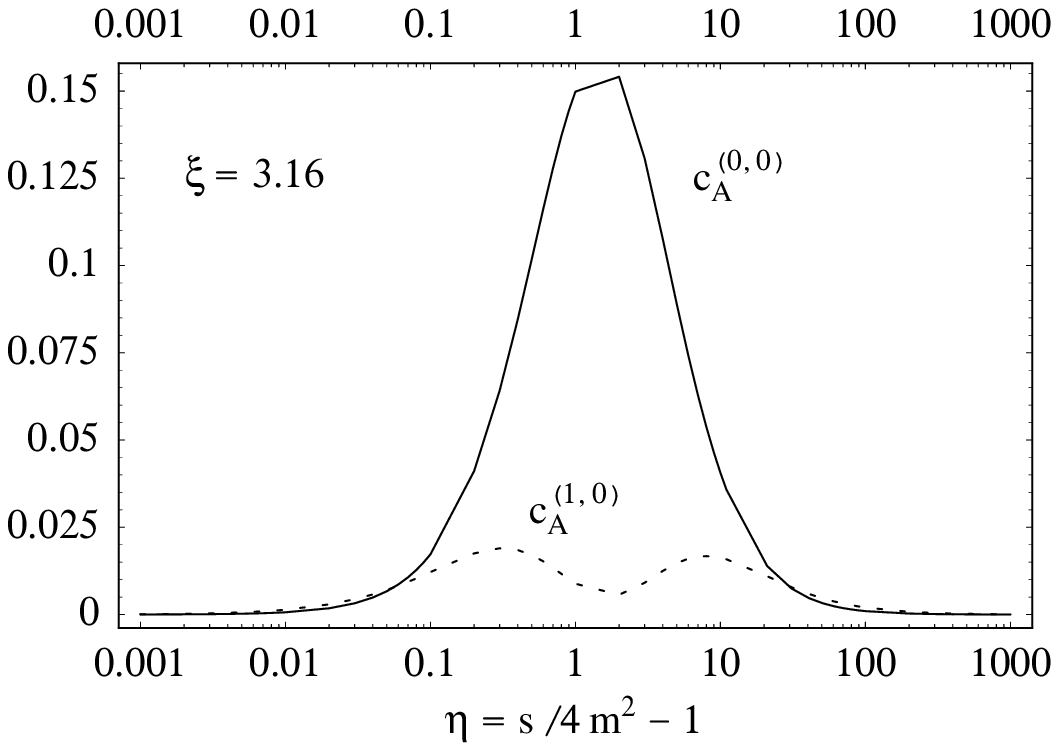,width=213pt}}&
\mbox{\epsfig{file=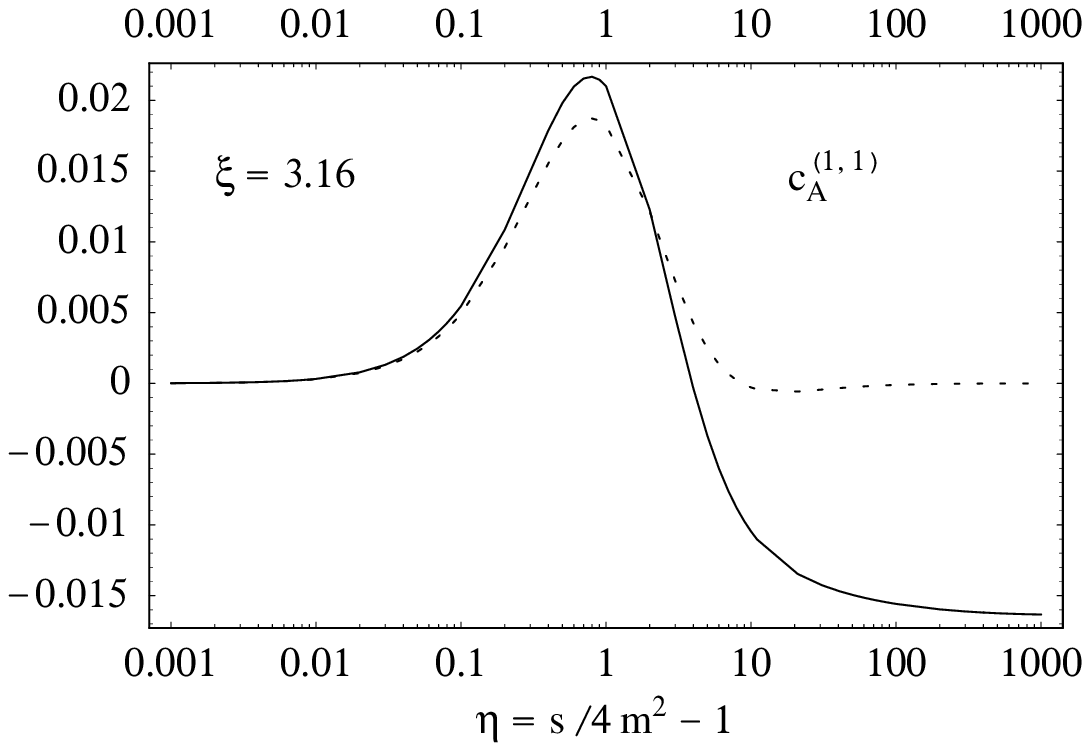,width=213pt}}\\
\end{tabular}
\caption{\label{Fg.4}\small $c_{T}^{(k,l)}(\eta,\xi )$ and $c_{A}^{(k,l)}(\eta,\xi )$
coefficient functions at $\xi =3.16$. Plotted are the available exact results
(solid lines) and the NLL approximation (dotted lines).}
\end{center}
\end{figure}

With Eq.~(\ref{25}) in hand, we are able to check the quality of the NLL
approximation against exact answers. In Figs.~\ref{Fg.3} and \ref{Fg.4} we
plot the functions $c_{T}^{(n,l)}(\eta ,\xi )$ and $c_{A}^{(n,l)}(\eta ,\xi )$
$(n,l=0,1)$ at $\xi =10^{-2}$ and $\xi =3.16$, respectively.
Predictions of the NLL approximation (\ref{20}) are given by dotted curves.
The available exact results are given by solid lines. One can see a
reasonable agreement up to energies $\eta \approx 2$. As to the $%
Q^{2}$-dependence, we have found that the soft-gluon approach reproduces
satisfactorily the exact results at $\xi \lesssim 1.$ At high values of $%
\xi $, $Q^{2}\gg m^{2}$, the quality of the NLL approximation becomes
worse.\footnote{%
Our analysis shows that the same situation takes also place for the energy
and $Q^{2}$ behavior of the functions $c_{L}^{(1,l)}(\eta ,\xi )$, $l=0,1$. We
do not give corresponding plots for $c_{L}^{(1,l)}(\eta ,\xi )$ because the
contribution of the longitudinal cross section to the $\cos 2\varphi $
asymmetry is small numerically.}

\section{Hadron Level Results}

\subsection{pQCD predictions}

Let us now analyze the impact of the approximate NLO perturbative
corrections on the AA at hadron level. We will consider the parameters $%
A(Q^{2})$, $A(x)$ and $A(y)$,
\[
A(Q^{2})=\frac{2\int\limits_{0}^{2\pi }\text{d}\varphi \cos 2\varphi
{\displaystyle {\text{d}^{2}\sigma _{lN} \over \text{d}Q^{2}\text{d}\varphi }}%
( \rho ,Q^{2},\varphi ) }{\int\limits_{0}^{2\pi }\text{d}\varphi
{\displaystyle {\text{d}^{2}\sigma _{lN} \over \text{d}Q^{2}\text{d}\varphi }}%
( \rho ,Q^{2},\varphi ) },\qquad \qquad A(x)=\frac{%
2\int\limits_{0}^{2\pi }\text{d}\varphi \cos 2\varphi
{\displaystyle {\text{d}^{2}\sigma _{lN} \over \text{d}x\text{d}\varphi }}%
( \rho ,x,\varphi) }{\int\limits_{0}^{2\pi }\text{d}\varphi
{\displaystyle {\text{d}^{2}\sigma _{lN} \over \text{d}x\text{d}\varphi }}%
( \rho ,x,\varphi ) },
\]
\begin{equation}
A(y)=\frac{2\int\limits_{0}^{2\pi }\text{d}\varphi \cos 2\varphi
{\displaystyle {\text{d}^{2}\sigma _{lN} \over \text{d}y\text{d}\varphi }}%
( \rho ,y,\varphi) }{\int\limits_{0}^{2\pi }\text{d}\varphi
{\displaystyle {\text{d}^{2}\sigma _{lN} \over \text{d}y\text{d}\varphi }}%
( \rho ,y,\varphi) },  \label{27}
\end{equation}
which describe the dependence of the $\cos 2\varphi $ asymmetry on $Q^{2}$,
Bjorken $x$ and $y$, respectively. Unless otherwise stated, the CTEQ5M \cite{28}
parametrization of the gluon distribution function is used. The default
values of the charm and bottom mass are $m_{c}=$ 1.5 GeV and $m_{b}=$ 4.75
GeV, $\Lambda _{3}=$ 260 MeV and $\Lambda _{4}=$ 200 MeV. For the
factorization scale we use $\mu _{F}=\sqrt{%
m_{b}^{2}+Q^{2}/4}$  in the case of bottom production and
$\mu _{F}=2\sqrt{m_{c}^{2}+Q^{2}/4}$ in the charm case \cite{1}.

\begin{figure}
\begin{center}
\begin{tabular}{cc}
\mbox{\epsfig{file=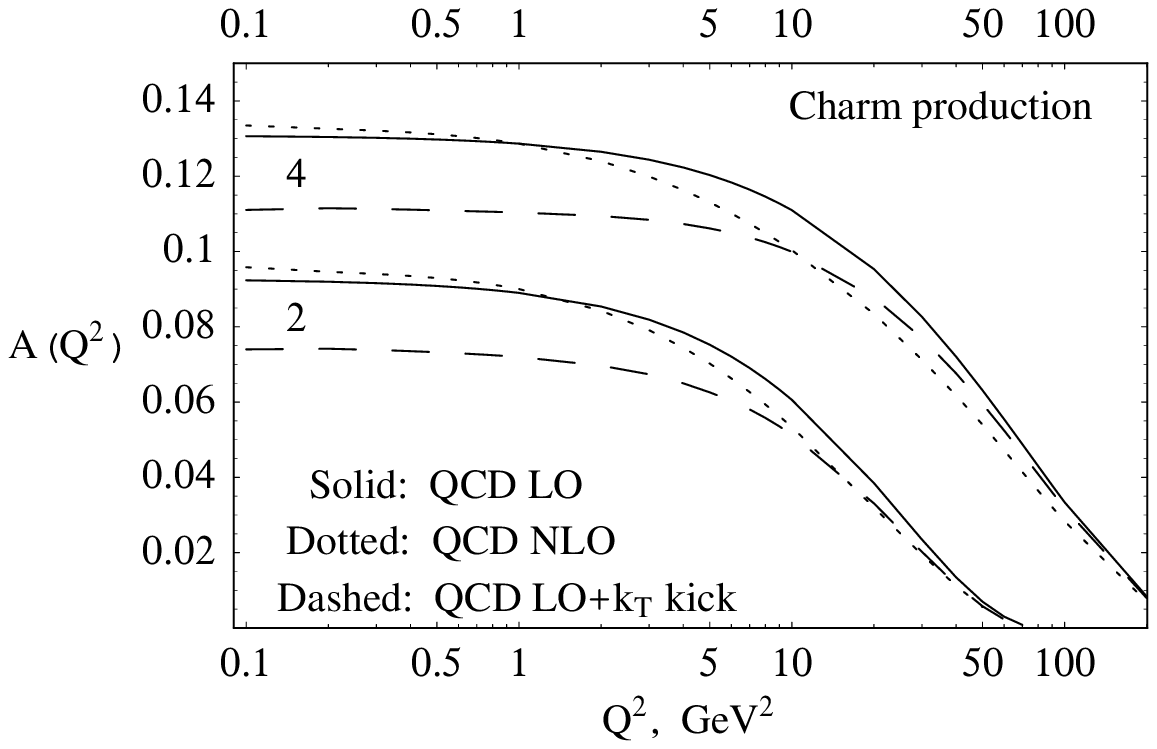,width=215pt}}&
\mbox{\epsfig{file=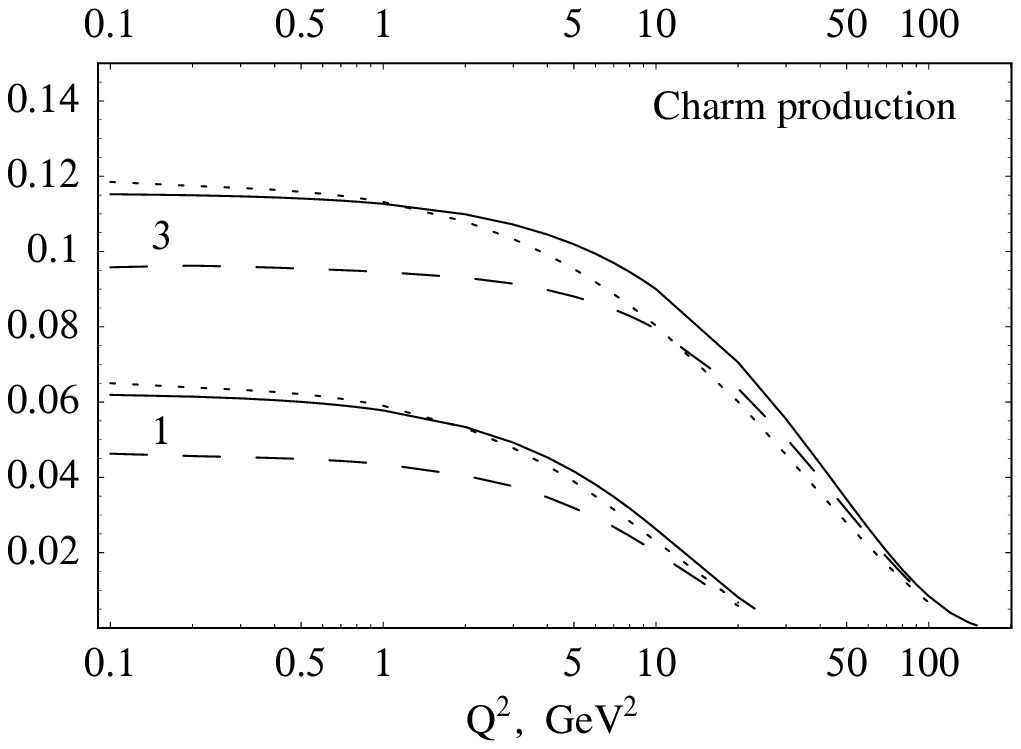,width=215pt}}\\
\end{tabular}
\caption{\label{Fg.5}\small Azimuthal asymmetry, $A(Q^{2})$, in $c$-quark
production for several values of initial energy: $E_{1}=24$ GeV, $E_{2}=47$ GeV,
$E_{3}=95$ GeV and $E_{4}=190$ GeV. Plotted are the results at LO (solid curve),
at NLO to NLL accuracy (dotted curve) and at LO with $k_{T}$ smearing (dashed curve).}
\end{center}
\end{figure}

\begin{figure}
\begin{center}
\begin{tabular}{cc}
\mbox{\epsfig{file=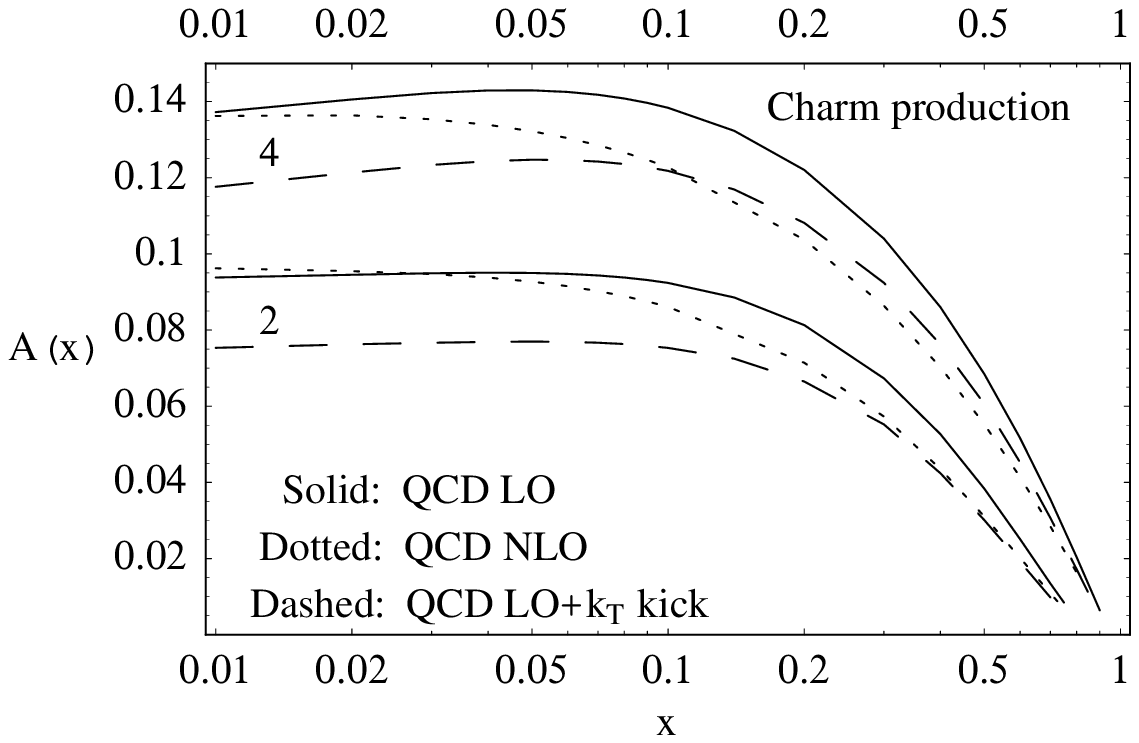,width=215pt}}&
\mbox{\epsfig{file=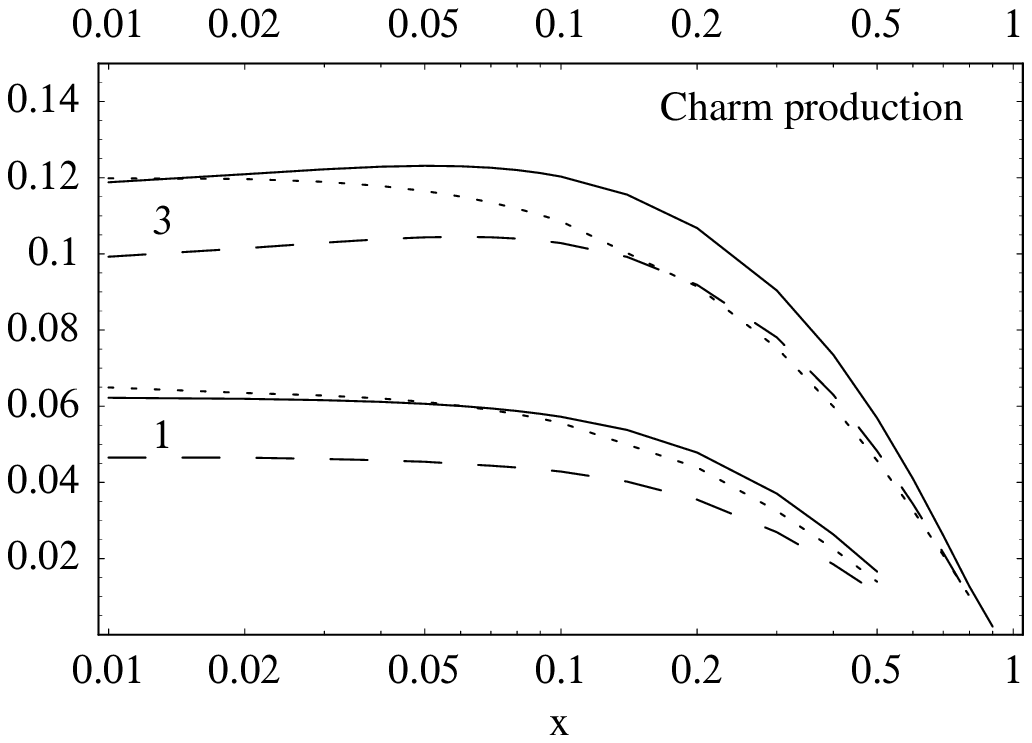,width=215pt}}\\
\end{tabular}
\caption{\label{Fg.6}\small Azimuthal asymmetry, $A(x)$, in $c$-quark production.
The notation is the same as in Fig.~\ref{Fg.5}.}
\end{center}
\end{figure}

\begin{figure}
\begin{center}
\begin{tabular}{cc}
\mbox{\epsfig{file=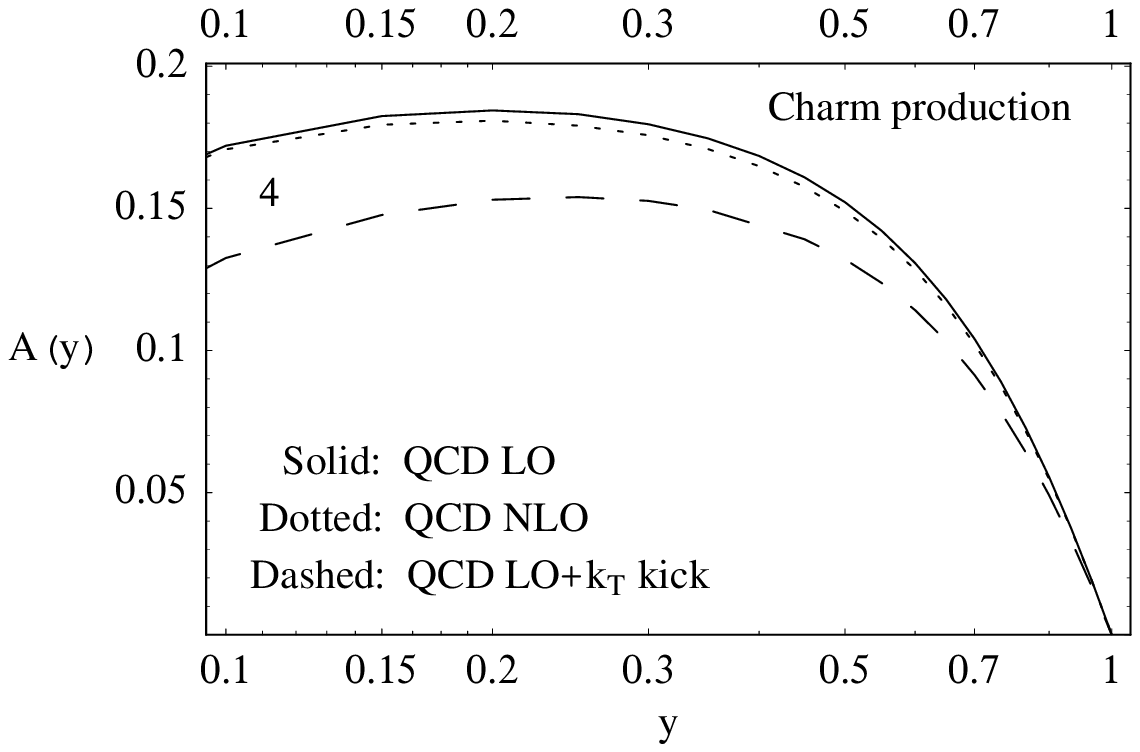,width=215pt}}&
\mbox{\epsfig{file=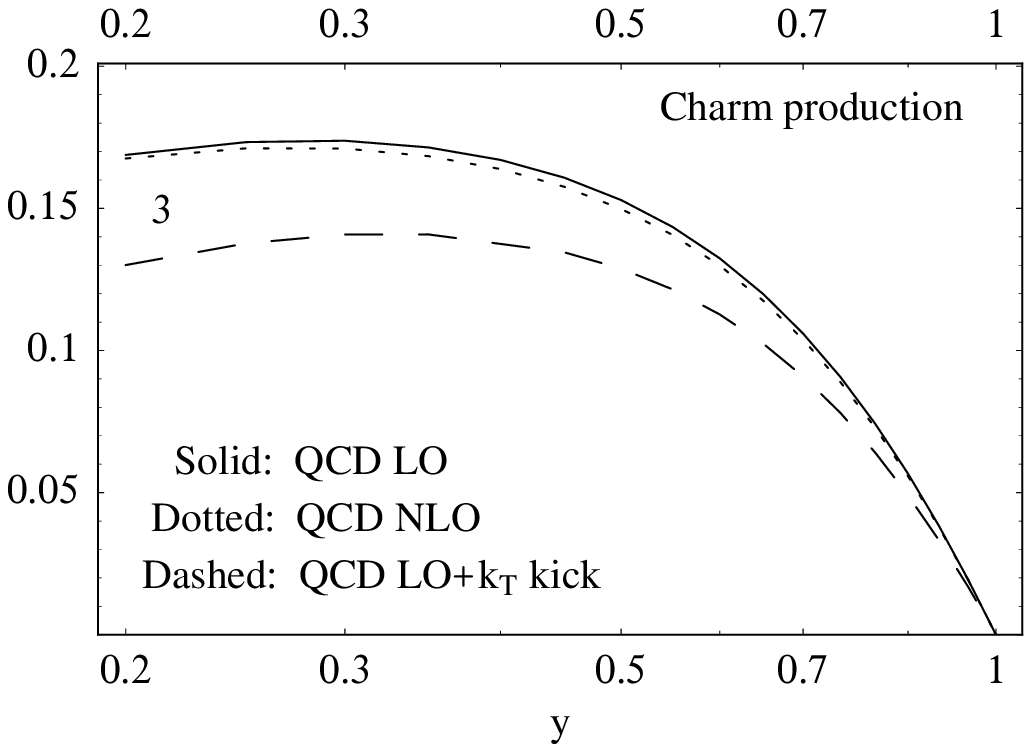,width=215pt}}\\
\end{tabular}
\caption{\label{Fg.7}\small Azimuthal asymmetry, $A(y)$, in $c$-quark production.
The notation is the same as in Fig.~\ref{Fg.5}.}
\end{center}
\end{figure}

\begin{figure}
\begin{center}
\begin{tabular}{cc}
\mbox{\epsfig{file=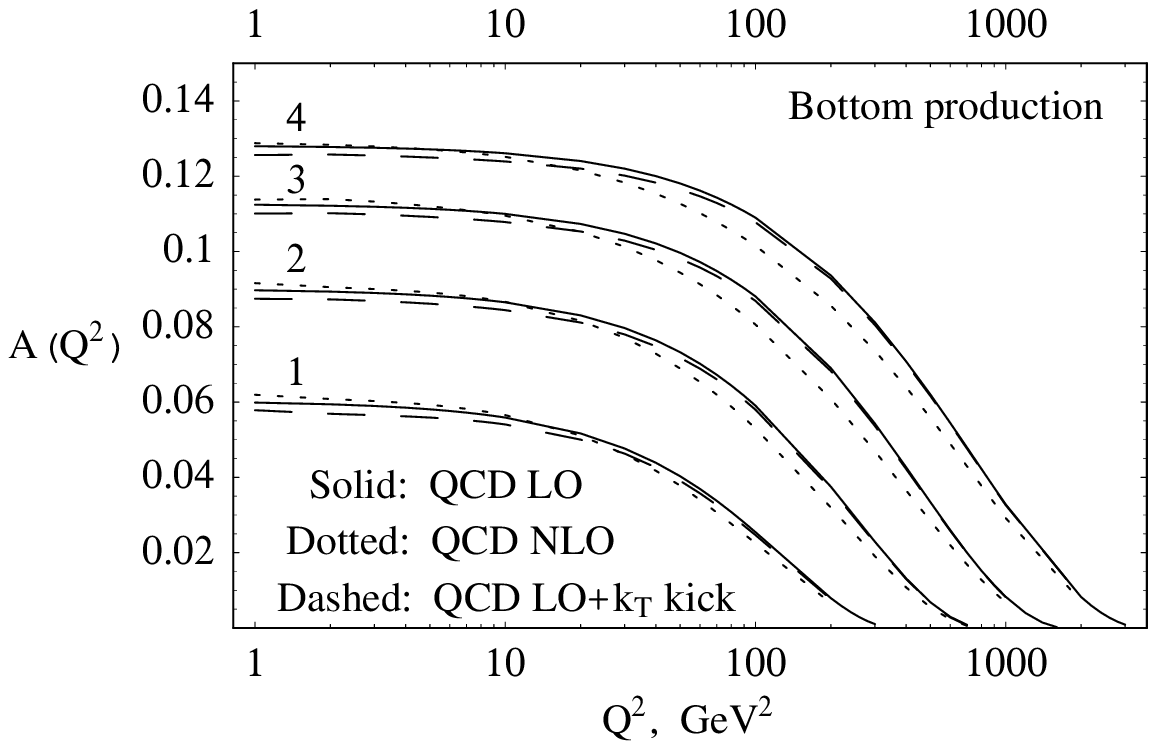,width=215pt}}&
\mbox{\epsfig{file=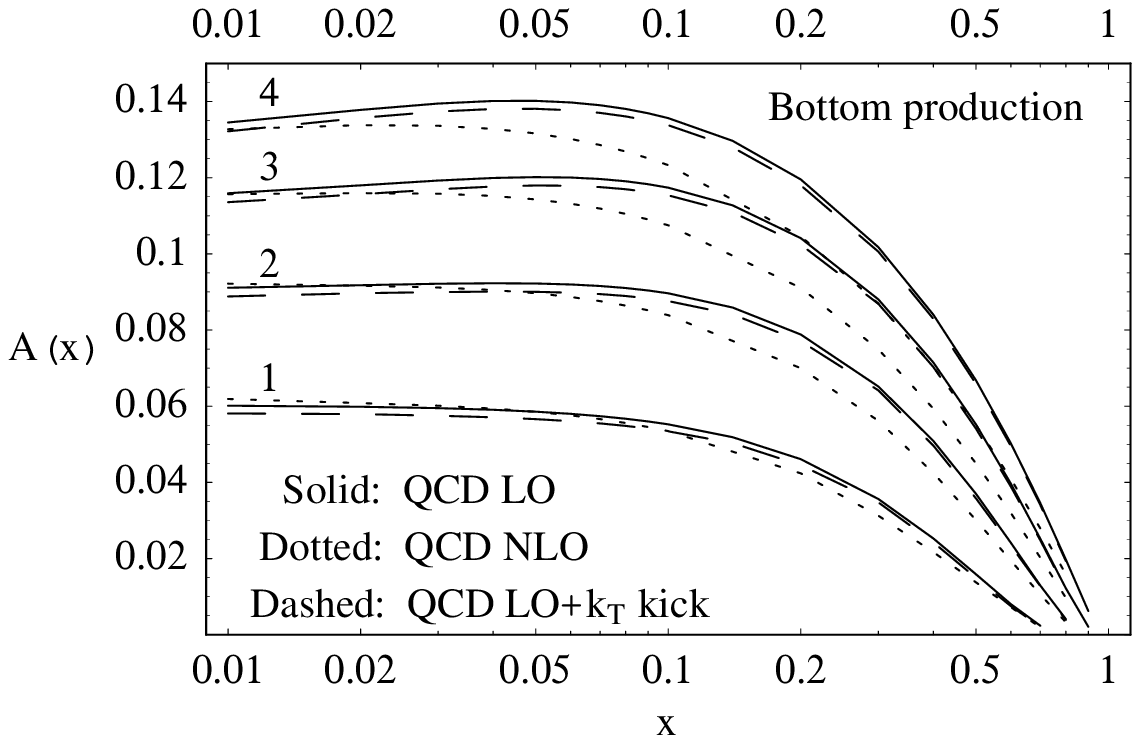,width=215pt}}\\
\multicolumn{2}{c}{\mbox{\epsfig{file=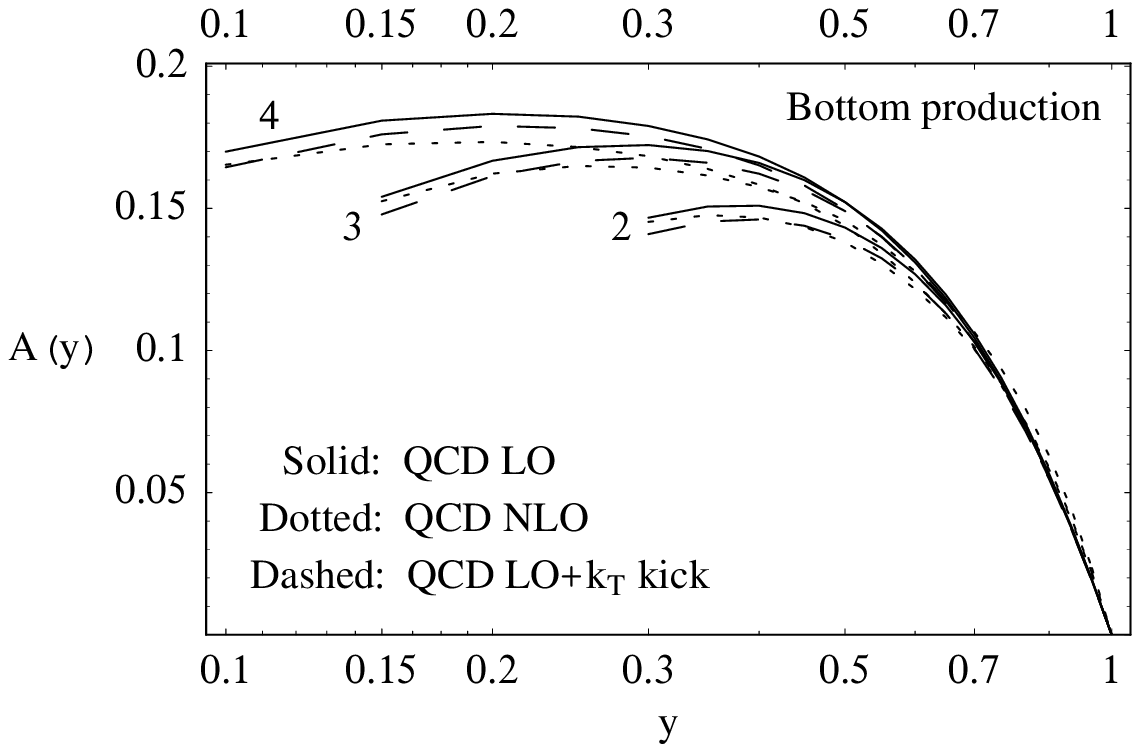,width=215pt}}}
\end{tabular}
\caption{\label{Fg.8}\small Asymmetry parameters $A(Q^{2})$ (left panel), $A(x)$ 
(right panel) and $A(y)$ (down panel) in $b$-quark production for several
values of initial energy: $E_{1}=240$ GeV, $E_{2}=480$ GeV, $E_{3}=960$ GeV and
$E_{4}=1920$ GeV. Plotted are the results at LO (solid curve), at NLO to NLL
accuracy (dotted curve) and at LO with $k_{T}$ smearing (dashed curve).}
\end{center}
\end{figure}

Our results for the $Q^{2}$- and $x$-distributions of the AA in charm
leptoproduction at several values of initial energy are presented in Fig.~\ref{Fg.5} and
Fig.~\ref{Fg.6}, respectively. The LO and NLO predictions are given by solid and
dotted lines, correspondingly. The lines with label $``1"$ correspond to $\rho _{1}=0.2$,
$``2"\rightarrow \rho _{2}=0.1$, $``3"\rightarrow \rho _{3}=0.05$ and
$``4"\rightarrow \rho _{4}=0.025$, where $\rho =4m^{2}/\bar{S}$. So, in the charm case,
we have: $E_{1}=24$ GeV, $E_{2}=47$ GeV, $E_{3}=95$ GeV and $E_{4}=190$ GeV,
where $E$ is the lepton energy in the lab (nucleon rest) frame:
$E=(\bar{S}-m_{N}^{2})/(2m_{N})$.  The $y$-distribution of the asymmetry at the
COMPASS energies is given in Fig.~\ref{Fg.7}.
In Fig.~\ref{Fg.8}  we plot the $A(Q^{2})$, $A(x)$ and $A(y)$
distributions of the asymmetry in bottom production at the same values of
$\rho _{i}$ $=\{0.2,0.1,0.05,0.025\}$ which correspond to the following set of initial
energies: $E_{i}=\{240,480,960,1920\}$ in units of GeV.

\begin{figure}
\begin{center}
\begin{tabular}{cc}
\mbox{\epsfig{file=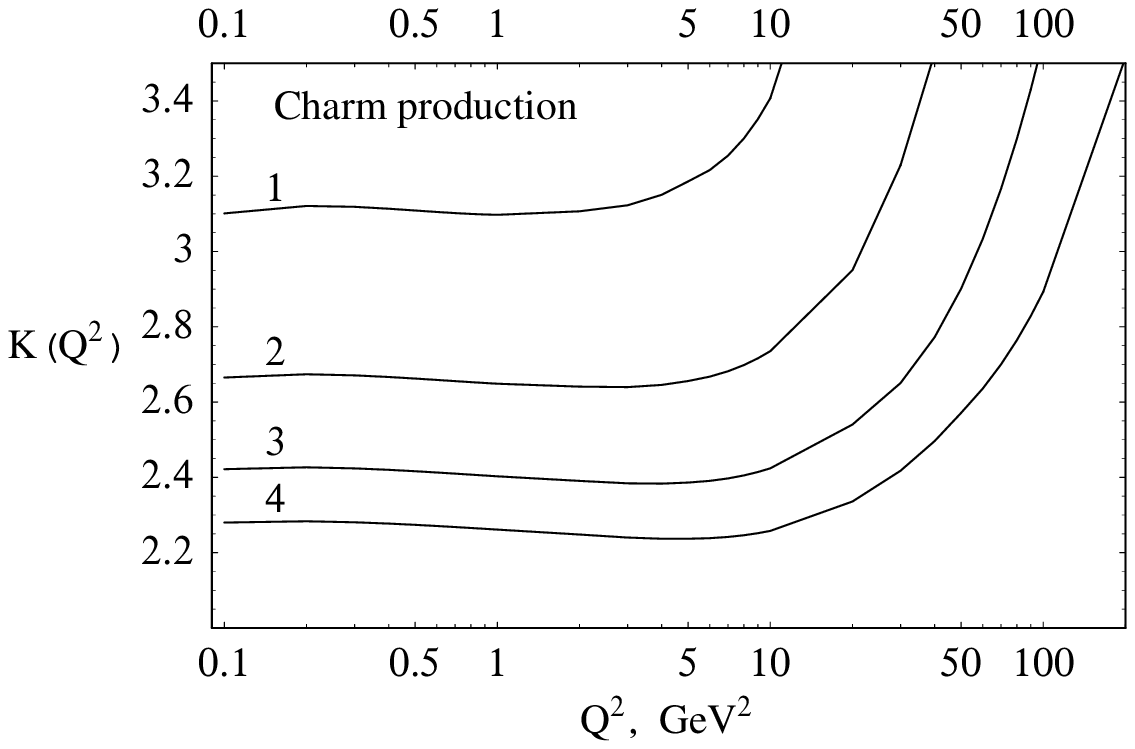,width=215pt}}
&\mbox{\epsfig{file=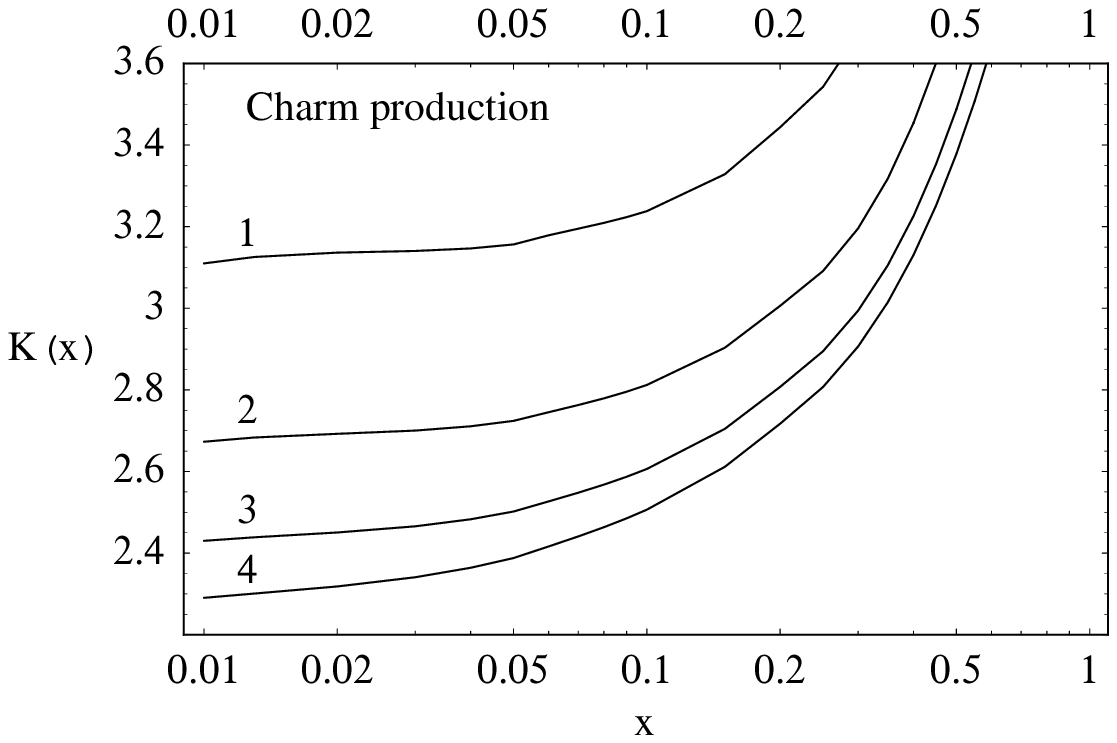,width=215pt}}\\
\end{tabular}
\caption{\label{Fg.10}\small $K$-factors in charm production at NLL level:
$K(Q^{2})=\left. \frac{\text{d}\sigma _{lN}^{{\rm NLO}}}{\text{d}%
Q^{2}}\right/ \frac{\text{d}\sigma _{lN}^{{\rm LO}}}{\text{d}Q^{2}}$ (left panel)
and $K(x)=\left. \frac{\text{d}\sigma _{lN}^{{\rm NLO}}}{\text{d}x}\right/ \frac{%
\text{d}\sigma _{lN}^{{\rm LO}}}{\text{d}x}$ (right panel).
The notation is the same as in Fig.~\ref{Fg.5}.}
\end{center}
\end{figure}

Our calculations given in Figs.~\ref{Fg.5}$-$\ref{Fg.8} represent the central
result of this paper. One can see from Figs.~\ref{Fg.5} and \ref{Fg.8} that
soft-gluon corrections to $A(Q^{2})$ are about few percent at not large
$Q^{2}\lesssim m^{2}$. At fixed values of $x$, the kinematical restriction
$Q^{2}\lesssim m^{2}$ leads to $x \lesssim m^{2}/\bar{S}$.
For this reason, radiative correction to $A(x)$ are small in the region of $x \lesssim \rho$
(see Figs.~\ref{Fg.6} and \ref{Fg.8}). For comparison, we plot in
Fig.~\ref{Fg.10} the so-called $K$-factors for $\varphi$-integrated cross
sections: $K(Q^{2})=\left. \frac{\text{d}\sigma _{lN}^{{\rm NLO}}}{\text{d}%
Q^{2}}\right/ \frac{\text{d}\sigma _{lN}^{{\rm LO}}}{\text{d}Q^{2}}$ and $%
K(x)=\left. \frac{\text{d}\sigma _{lN}^{{\rm NLO}}}{\text{d}x}\right/ \frac{%
\text{d}\sigma _{lN}^{{\rm LO}}}{\text{d}x}$. One can see that large soft-gluon
corrections to the production cross sections practically (to within few percent) do not
affect the Born predictions for the $\cos 2\varphi $ asymmetry at $Q^{2}\lesssim m^{2}$
and $x \lesssim \rho$.

At fixed values of $y$, the allowed region of $Q^{2}$ is
$m^{2}_{l}y^{2}/(1-y)\le Q^{2}\le y\bar{S}-4 m^{2}$, where $m_{l}$ is the initial
lepton mass. Since production cross sections rapidly vanish with growth of $Q^{2}$,
practically whole contribution to $A(y)$ originates from the low-$Q^{2}$ region.
For this reason, radiative corrections to $A(y)$ are negligible
practically in the whole region of $y$ (see Figs.~\ref{Fg.7} and \ref{Fg.8}).

 Let us now discuss the region of applicability of the adopted
soft-gluon approximation. As noted in previous Section, soft radiation
reproduces satisfactorily the existing exact NLO results when $\xi $ is not
large, $Q^{2}\lesssim m^{2}$. At large $Q^{2}\gg m^{2}$, the hard $(\vec{p}%
_{g,T}\neq 0)$ contributions becomes sizeable and the quality of the NLL
approximation becomes worse for both $\varphi $-dependent and $\varphi $%
-independent cross sections. Moreover, we have observed that soft-gluon
approximation overestimates the exact results for $c_{T}^{(1,0)}(\eta ,\xi )$
and $c_{T}^{(1,1)}(\eta ,\xi )$ at $\xi \gg 1$ and, simultaneously,
underestimates the corresponding ones for $c_{A}^{(1,1)}(\eta ,\xi )$. For
this reason, in the high-$Q^{2}$ region, the full NLO corrections to the $%
\cos 2\varphi $ asymmetry may be smaller than the soft-gluon ones.

As to the energy dependence,
one can see from Figs.~\ref{Fg.3} and \ref{Fg.4} that soft
radiation describes very well the exact NLO results on $\varphi $-independent
photon-gluon fusion at partonic energies up to $\eta \approx 2$. Since the
gluon distribution function supports just the threshold region, the
soft-gluon contribution dominates the photon-hadron cross sections
approximately up to $S/4m^{2}\sim 10$ (and, correspondingly, up to $\rho
\sim 0.1$ for $\sigma _{lN}$). Using the exact results for the $\gamma
^{*}g $ cross sections \cite{26}, we have verified that the contribution
originating from the region $\eta >2$ makes only few percent from the NLO
hadron-level predictions for $\sigma _{T}(S,Q^{2})$ and $\sigma
_{L}(S,Q^{2}) $ at $S/4m^{2}\lesssim 10$ (and $Q^{2}\lesssim m^{2}$).
Results of Ref.\cite{24} on soft-gluon corrections to $F_{2}^{{\rm charm}%
}(x,\xi )$ confirm our conclusion.

Presently, exact NLO calculations of the $\varphi $-dependent cross
section of heavy flavor production are not completed. However we can be sure
that, at energies not so far from the production threshold, the soft
radiation is the dominant perturbative mechanism in the case of $\sigma
_{A}(S,Q^{2})$ too. First, LO predictions for the $\cos 2\varphi $%
-dependent cross section are large and the Sudakov logarithms have
universal, $\varphi $-independent structure. For this reason, $\sigma
_{A}(S,Q^{2})$ has also a strong threshold enhancement. Second, our analysis
of the exact scale-dependent cross section $c_{A}^{(1,1)}(\eta ,\xi )$ given
in Figs.~\ref{Fg.3} and \ref{Fg.4} confirms with a good accuracy the dominance
of the soft-gluon contribution. These facts argue that hard and virtual corrections
to the $\cos 2\varphi $-dependent cross section cannot affect significantly the
soft-gluon predictions for the azimuthal asymmetry at low $Q^{2}$ in the
energy region up to $S/4m^{2}\sim 10$.
\begin{figure}
\begin{center}
\begin{tabular}{cc}
\mbox{\epsfig{file=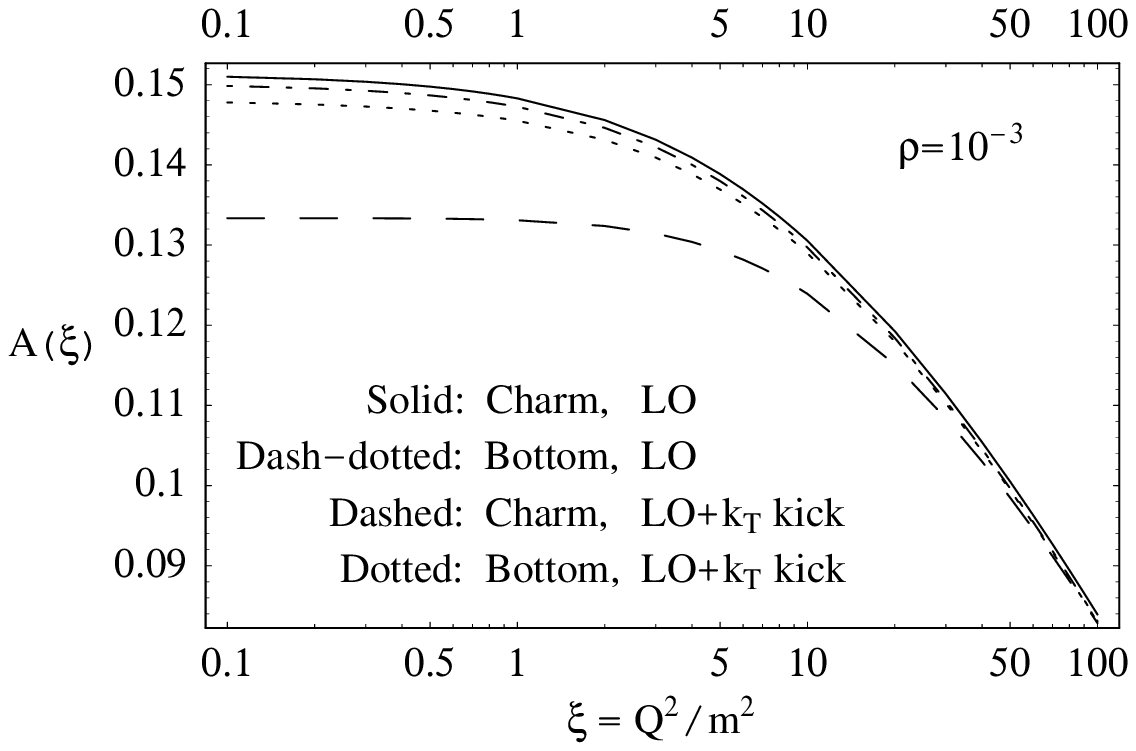,width=215pt}}
&\mbox{\epsfig{file=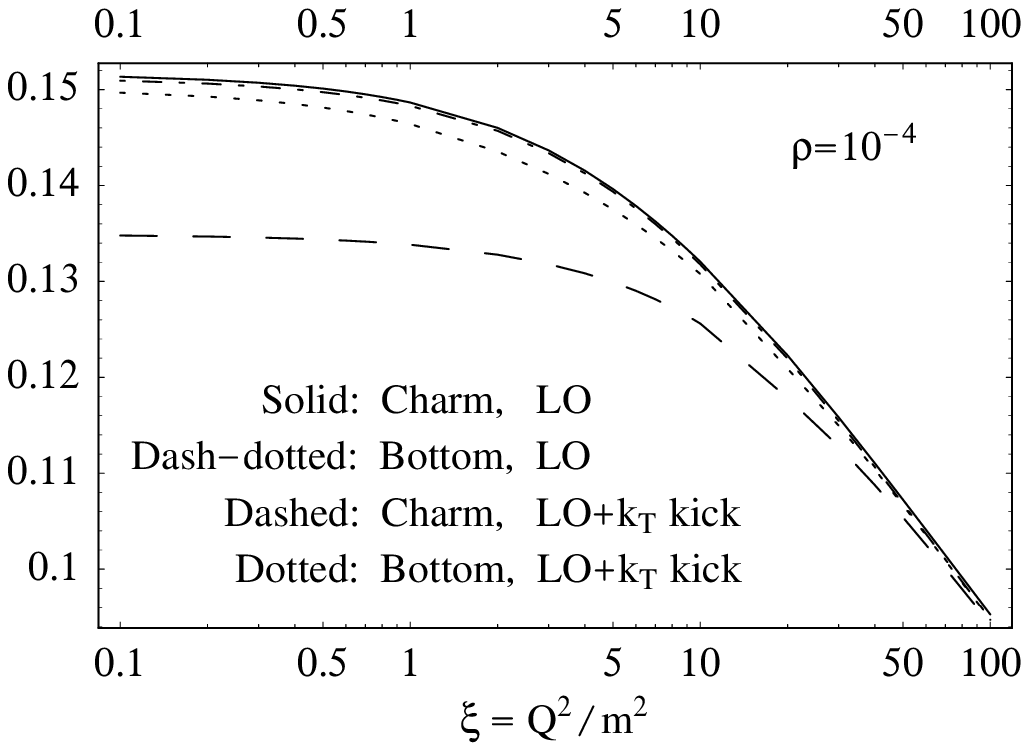,width=215pt}}\\
\end{tabular}
\caption{\label{Fg.12}\small QCD LO predictions for $A(\xi)$ in charm and bottom 
production at $\rho=10^{-3}$ (left panel) and $\rho=10^{-4}$ (right panel) 
with and without the inclusion of the $k_{T}$ smearing effect.}
\end{center}
\end{figure}
Let us briefly discuss the origin of perturbative stability of the $\cos
2\varphi $ asymmetry. Note that the mere $\varphi $-independent structure of
the Sudakov logarithms cannot explain our results since perturbative
stability does not take place at the parton level. In fact, the ratios $%
\frac{c_{A}^{(1,0)}}{c_{T}^{(1,0)}}(\eta ,\xi )$ and $\frac{c_{A}^{(0,0)}}{%
c_{T}^{(0,0)}}(\eta ,\xi )$ differ essentially from each other even at $\eta
,\xi \lesssim 1$. This is due to the fact that the physical soft-gluon
corrections (\ref{18}) are determined by a convolution of the Born cross
sections with the Sudakov logarithms which, apart from factorized $\delta
(s_{4})$-terms, contain also non-factorizable ones (see Eq. (\ref{22})).
Kinematically, sizeable values of $\eta \sim 1$ allow $s_{4}/m^{2}\sim 1$
that leads to significant non-factorizable corrections. In other words,
collinear bremsstrahlung carries away a large part of initial energy. Since
the $\varphi $-dependent and $\varphi $-independent Born level partonic
cross sections have different energy behavior, the soft radiation has
different impact on these quantities.

Our analysis shows that two more factors are responsible for perturbative
stability of the hadron level asymmetry. First, one can see from
Figs.~\ref{Fg.3} and \ref{Fg.4}
that both $\varphi $-dependent and $\varphi $-independent Born level cross
sections take their maximum values practically at the same values of $\eta$.
Second, at fixed target energies, the gluon distribution function supports the
contribution of the threshold region. In other words, sufficiently soft
gluon distribution function makes the collinear gluon radiation effectively
soft at the hadron level. In detail, the role of the gluon distribution
function in perturbative stability of the azimuthal asymmetry in heavy quark
photoproduction is discussed in Ref.\cite{14}.

Another remarkable property of the azimuthal asymmetry closely related to
fast perturbative convergence is its parametric stability.\footnote{Of course,
parametric stability of the fixed order results does not imply a
fast convergence of the corresponding series. However, a fast convergent
series must be parametrically stable. In particular, it must be $\mu _{R}$-
and $\mu _{F}$-independent.} Our analysis shows that the pQCD predictions
for the $\cos 2\varphi $ asymmetry are less sensitive to standard uncertainties in
the QCD input parameters than the corresponding ones for the production cross
sections.  For instance, changes of $\mu _{F}$ in the range $m_{c}<\mu _{F}<2%
\sqrt{m_{c}^{2}+Q^{2}/4}$ affect the quantity $A(Q^{2})$ in charm production
by less than $7\%$ at $\rho =0.025$ and $\xi \leq 4$. For the $\varphi $%
-integrated cross section, such changes of $\mu _{F}$ lead to $30\%$
variations in the same kinematics. We have also verified that all the NLO
CTEQ5 versions of gluon density as well as the LO parametrization \cite
{28a} lead to asymmetry predictions which coincide with each other with an
accuracy of better than $1.5\%$.

Parametric stability of the azimuthal asymmetry leads to the scaling: with a good
accuracy the quantity $A(\rho ,x,\xi )$ in Eq. (\ref{6}) is a function of three
variables, so that
\begin{equation}
A^{{\rm {Charm}}}(\rho ,x,\xi )\approx A^{{\rm {Bottom}}}(\rho ,x,\xi )
\label{A1}
\end{equation}
at the same values of $\rho $, $x$ and $\xi $. To illustrate this property, 
in Fig.~\ref{Fg.12} we plot the asymmetry parameter $A(Q^2)$ defined by 
Eqs.~(\ref{27}) as a function of $\xi$ at $\rho=10^{-3}$ and $10^{-4}$.  
Since the soft-gluon approximation 
is inapplicable to heavy flavor production at high energies, we give only the 
LO predictions for $A(\xi)$. The LO results for the charm and bottom cases are 
plotted by solid and dash-dotted lines, respectively. One can see that both 
curves coincide with each other with an accuracy of better than $1\%$. 

It is also seen from Fig.~\ref{Fg.12} that, at the HERA energy 
($\rho \sim 10^{-3}$ and  $10^{-4}$ for bottom and charm quark, respectively), 
the pQCD predictions for the AA in heavy quark leptoproduction are large 
and can be tested experimentally. 

\subsection{Nonperturbative corrections}

Let us discuss how the pQCD predictions for azimuthal asymmetry are affected
by nonperturbative contributions due to the intrinsic transverse motion of
the gluon in the target. Because of the relatively low $c$-quark mass, these
contributions are especially important in the description of the cross
sections for charmed particle production \cite{1}.

To introduce $k_{T}$ degrees of freedom, $\vec{k}_{g}\simeq z\vec{p}+\vec{k}%
_{T}$, one extends the integral over the parton distribution function in Eq.
(\ref{13}) to $k_{T}$-space, 
\begin{equation}
\text{d}zg(z,\mu _{F})\rightarrow \text{d}z\text{d}^{2}k_{T}f\big( \vec{k}%
_{T}\big) g(z,\mu _{F}).  \label{28}
\end{equation}
The transverse momentum distribution, $f\big( \vec{k}_{T}\big) $, is
usually taken to be a Gaussian: 
\begin{equation}
f\big( \vec{k}_{T}\big) =\frac{{\rm {e}}^{-k_{T}^{2}/\langle
k_{T}^{2}\rangle }}{\pi \langle k_{T}^{2}\rangle }.  \label{29}
\end{equation}
In practice, an analytic treatment of $k_{T}$ effects is usually used.
According to \cite{29}, the $k_{T}$-smeared differential cross section of
the process (\ref{1}) is a 2-dimensional convolution: 
\begin{equation}
\frac{\text{d}^{4}\sigma _{lN}^{{\rm {kick}}}}{\text{d}x\text{d}Q^{2}\text{d}%
p_{QT}\text{d}\varphi }\left( \vec{p}_{QT}\right) =\int \text{d}^{2}k_{T}%
\frac{{\rm {e}}^{-k_{T}^{2}/\langle k_{T}^{2}\rangle }}{\pi \langle
k_{T}^{2}\rangle }\frac{\text{d}^{4}\sigma _{lN}}{\text{d}x\text{d}Q^{2}%
\text{d}p_{QT}\text{d}\varphi }\Big( \vec{p}_{QT}-\frac{1}{2}\vec{k}%
_{T}\Big) .  \label{30}
\end{equation}
The factor $\frac{1}{2}$ in front of $\vec{k}_{T}$ in the r.h.s. of Eq. (\ref
{30}) reflects the fact that the heavy quark carries away about one half of
the initial energy in the reaction (\ref{1}).

Values of the $k_{T}$-kick corrections to LO predictions for the $\cos 2\varphi $ 
asymmetry in the charm production are shown in Figs.~\ref{Fg.5}$-$\ref{Fg.7} 
and \ref{Fg.12} by dashed curves. Calculating the $k_{T}$-kick effects we use 
$\langle k_{T}^{2}\rangle =0.5$ GeV$^{\text{2}}$. At fixed target energies, 
$k_{T}$-smearing for $A(Q^{2})$ and $A(x)$ is about $20$-$25\%$ in the 
region of low $Q^{2}$ and $x$, respectively, and decreases at large $Q^{2}$ 
and $x$.  In the HERA range, expected values of the $k_{T}$-corrections are 
systematically smaller (see Fig.~\ref{Fg.12}).

Analogous calculations for the case of bottom production are presented
in Figs.~\ref{Fg.8} and \ref{Fg.12}. It is seen that $k_{T}$-kick 
corrections to the $b$-quark AA are practically negligible in the whole region 
of $Q^{2}$ and $x$.

\section{Conclusion}

In this paper we have investigated the impact of soft-gluon radiation on
the $\cos 2\varphi $ asymmetry in heavy flavor leptoproduction. 
The NLL approximation provides a good description of the available
exact NLO results for $Q^{2}\lesssim m^{2}$ at energies of the fixed target
experiments.  Our calculations show that the azimuthal asymmetry is
practically insensitive to soft-gluon corrections in this kinematics. 
We conclude that, unlike the $\varphi $-integrated cross sections, 
the $\cos 2\varphi $ asymmetry in heavy quark leptoproduction
is an observable quantitatively well defined in pQCD: it is stable both
parametrically and perturbatively, and insensitive (in the case of bottom
production) to nonperturbative contributions. This asymmetry is of leading
twist and predicted to be about $15\%$ at energies sufficiently above the 
production threshold for both charm and bottom quark. 
Measurements of the $\cos 2\varphi $ asymmetry in bottom production at 
HERA would provide an ideal test of pQCD. 

Data on the charm azimuthal distributions from the COMPASS and 
HERMES experiments would make it possible to clarify the role of subleading 
twist contributions.  Our analysis shows that, in the low-$x$ region, the AA is 
sensitive to the gluon transverse motion in the target. At high $x$,  the intrinsic 
charm contribution \cite{30} to the structure functions may be significant \cite{31,32}. 
In detail, the possibility of measuring the intrinsic charm content of the proton 
using the $\cos 2\varphi $ asymmetry will be considered in a forthcoming publication. 

\acknowledgements 
\noindent The author would like to thank S.J. Brodsky, A.B. Kaidalov, A.Kotzinian 
and A.G. Oganesian for useful discussions. I am grateful to High Energy Section 
of ICTP for hospitality while this work has been completed. 
\def\baselinestretch{1.}

\end{document}